\documentclass[aps,prb,twocolumn,groupedaddress,amsmath,amssymb,amsfonts,graphics]{revtex4}
\usepackage{feynmf}
\usepackage{bm}
\usepackage{wasysym}
\usepackage{ulem}
\usepackage{graphicx}
\usepackage{bm}
\usepackage{color}
\usepackage{verbatim}
\usepackage{slashed}

\def\l{\left}
\def\r{\right}
\def\d{^{\dagger}}
\def\<{\begin{equation}}
\def\>{\end{equation}}
\newcommand{\ket}[1]{\left| #1\right\rangle}              % | >

\begin{document}

%\raggedbottom

%\title{Finite coupling Kondo insulator transition in the honeycomb lattice}

\title{Quantum critical point  in the Kondo--Heisenberg model on the honeycomb lattice}

\author{Saeed Saremi}
\affiliation{Department of Physics, Massachusetts Institute of Technology,
Cambridge, Massachusetts 02139}
\author{Patrick A.\  Lee}
\affiliation{Department of Physics, Massachusetts Institute of Technology,
Cambridge, Massachusetts 02139}

\date{Octobor 10, 2006}
%\date{\today}

\begin{abstract}
We study the Kondo--Heisenberg model on the honeycomb lattice at half-filling. Due to the vanishing of the density of states at the fermi level, the Kondo insulator disappears at a finite Kondo coupling  even in the absence of the Heisenberg exchange. We adopt a large-N formulation of this model and use the renormalization group machinery to study systematically the second order phase transition of the Kondo insulator (KI) to the algebraic spin liquid (ASL). We note that neither phase breaks any physical symmetry, so that the transition is not described by the standard Ginzburg-Landau-Wilson critical point. We find a stable Lorentz-invariant fixed point that controls this second order phase transition. We calculate the exponent $\nu$ of the diverging length scale near the transition. The quasi-particle weight of the conduction electron vanishes at this KI--ASL fixed point, indicating non-Fermi liquid behavior. The algebraic decay exponent of the staggered spin correlation is calculated at the fixed point and in the ASL phase. We find a jump in this exponent at the transition point.\end{abstract}

\maketitle

%\tableofcontents

%**********************************************
\section{Introduction \label{SEC:Introduction}}
%**********************************************
The interplay between charge and spin degrees of freedom has been a focus of research in complex materials such as cuprates and heavy-fermions.\cite{Coleman05} The clearest example of this interplay is the quantum critical point seen in many heavy-fermion materials. At zero temperature, the heavy Fermi liquid (HFL) phase disappears exactly at a point where the anti-ferromagnetic (AF) magnetic ordering grows. Non-Fermi liquid behaviors are seen in the quantum critical region, i.e. the V-shaped region above this critical point.\cite{Coleman05} The theoretical understanding of why these two seemingly different phases, i.e. AF ordered and heavy fermi liquid, should collapse at one point and a clear understanding of the non-Fermi liquid behaviors in the quantum critical region are poor at the moment. 

It is by  now understood that the spin density wave approach to understand the quantum critical point in heavy-fermions, known as the Moriya--Hertz--Millis theory \cite{MHM}, only accounts for small deviations from Fermi liquid theory. New theoretical approaches for understanding the quantum critical heavy-fermions are needed.  

In an attempt to find an alternative for the Moriya--Hertz--Millis theory, Senthil \textit{et al.} proposed recently that the AF--HFL transition might be controlled by an \textit{unstable} spin liquid fixed point. \cite{SenthilVojta03, SenthilVojta05} This picture is very similar to the Deconfined quantum critical point in the context of the second order phase transition between  Neel and  valence bond solid (VBS) ground states. \cite{Senthiletal}  In that case the transition to VBS  happens due to the existence of an unstable spin liquid on the magnetically disordered side of the critical point. Deconfined quantum critical points open up the possibility of second order phase transitions between ``unrelated" phases like AF and HFL. It also has the advantage of giving a plausible scenario for the non-Fermi liquid behavior seen in experiments in heavy fermion materials. This view has had some successes in explaining some of the experimental observations.\cite{Marston05, Pepin06} In this paper, we further explore this point of view.

In search for a microscopic model that provides this type of quantum critical point, we study the Kondo--Heisenberg model on the \textit{honeycomb} lattice at \textit{half-filling}. The heavy fermion quantum critical point in this model is simplified. It corresponds to a point, where both charge gap and spin gap vanishes. This model is given by the following Hamiltonian ($J_K>0, J_H>0$):
\begin{equation} \label{INT:Hamiltonian}
\hat{H} = -t \sum_{\left<ij\right>,a} \left( c_{i}^{a\dagger} c_{j}^{a} + H.c. \right) + J_K \sum_i \bm{s}_i\cdot \bm{S}_i + J_H \sum_{\left<ij\right>} \bm{S}_i\cdot \bm{S}_j,
\end{equation}
where $i$ and $j$ live on the honeycomb lattice sites and $a$ is the spin index: $\{\uparrow,\downarrow\}$. The $\bm{s}_i$ and $\bm{S}_i$ denote the conduction electron spin and the localized spin at the site $i$ respectively. What makes the honeycomb lattice very interesting to study is the fact that the Kondo gap vanishes at a \textit{finite} coupling constant even as $J_H\rightarrow0$. This is due to the fact that, in contrast to the square lattice, the density of states of the conduction electrons vanishes at the Fermi level.\cite{Fradkin96}

The Kondo--Heisenberg model on the honeycomb lattice at half-filling is also interesting, because it can be studied by quantum Monte Carlo \textit{without} the fermion sign problem.\cite{Assaad99, Beach04} The present paper can be considered a prelude to such a study. In this connection we mention the quantum Monte Carlo study of the Kondo lattice model ($J_H=0$) on the square lattice by Assaad.\cite{Assaad99} He found a quantum phase transition of the AF ordered ground state to a disordered one caused by the Kondo exchange. The quasi-particle gap however did not vanish at the transition point, \footnote{See Fig. 6 of the mentioned Assaad's paper.}  i.e. the magnetic transition happened inside the KI phase. This made the transition a magnetic transition belonging to the O(3) nonlinear sigma model universality class. Note that the nested Fermi line of the tight-binding band  provides the instability towards AF at  $(\pi,\pi)$ even in the absence of the Heisenberg exchange $J_H$. In contrast the honeycomb lattice tight binding band has Fermi points in the corner of the Brillouin zone called the Dirac points. This causes the nesting to be extremely weak as compared to the perfectly nested lines of the Fermi line in the square lattice.

Next we consider different scenarios for the ground state phase diagram of our model, where we consider varying $J_K$ ($t=1$), keeping $J_H$ to be small. %@@@
In the limit $J_K \ll J_H$, the Heisenberg exchange dominates and will lead to Neel ordering with opposite spins on the AB sublattices of the honeycomb lattice.

\begin{figure}%[h]
\includegraphics{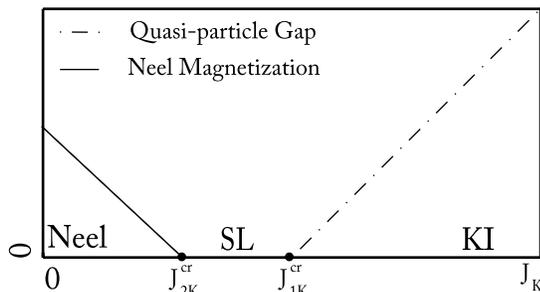}
\caption{ \label{FIG:KI_SL_Neel}
A scenario for the zero temperature phase diagram of the Kondo--Heisenberg Hamiltonian on the honeycomb lattice at half-filling [given by Hamiltonian Eq.~(\ref{INT:Hamiltonian})] as a function of $J_K$ ($t=1$) for a sufficiently small $J_H$.  In contrast to the same model on a square lattice,  Kondo gap denoted schematically by the dotted-dashed line vanishes at a finite coupling constant $J_{1K}^{cr}$. The weak nesting of the Dirac nodes of the honeycomb lattice makes the KI transition to a spin liquid ground state plausible.
}
\end{figure}

In the first scenario, sketched schematically in Fig.~\ref{FIG:KI_SL_Neel}, the finite coupling Kondo transition is to a magnetically disordered  ground state (spin liquid). In this scenario the AF ordered ground state does not survive up to the KI transition point for sufficiently small $J_H$. The Kondo physics has stabilized a spin liquid ground state in the region between the KI and AF ground states. The charge gap closes at $J_{1K}^{cr}$, as the KI gives way to a semi-metal. The spin gap may or may not be finite, depending on the nature of the spin liquid. As $J_K$ is further reduced a transition to the Neel phase occurs. This scenario will of course be very exciting, if a spin liquid can indeed be stabilized as the ground state over some parameter range.

\begin{figure}
\includegraphics{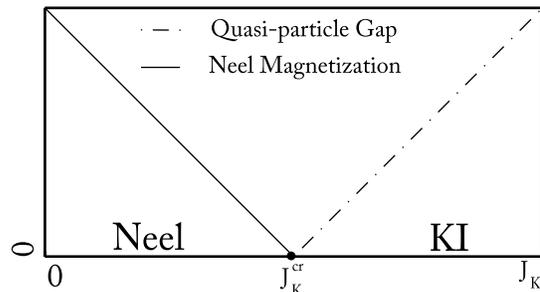}
\caption{ \label{FIG:KI_Neel}
Another scenario for the zero-tempreature phase diagram for the Kondo--Heisenberg Hamiltonian at half-filling on the honeycomb lattice for some (maybe ``fine tuned") Heisenberg exchange $J_H$ as a function of $J_K$. In contrast to the same model on a square lattice, both the quasi-particle gap denoted schematically by the dotted-dashed line and Neel order paramater denoted by the solid line vanishes at $J_K=J_K^{cr}$.
}
\end{figure}

The other exciting possibility is that a continuous transition between two distinct orders (AF and KI) exists over a finite parameter range. This scenario is sketched schematically in Fig.~\ref{FIG:KI_Neel}. In this scenario SL is unstable to Neel ordering. This will be an example of the deconfined quantum critical point  mentioned earlier. Finally, scenarios in which an overlapping region, where a Neel state coexists with the Kondo insulator (similar to the square lattice result of Assaad \cite{Assaad99}); or a first order transition can not be ruled out. With these motivations we now sketch an outline of this paper.

\begin{figure}[h]
\includegraphics{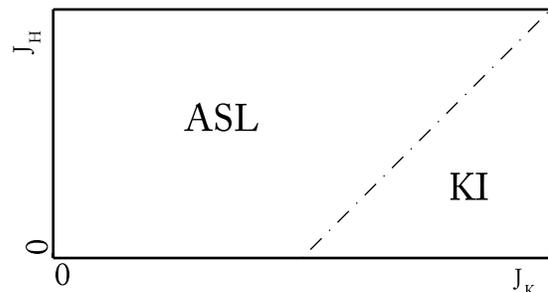}
\caption{ \label{FIG:KI_ASL}
The mean-field phase diagram of the Hamiltonian given by Eq.~(\ref{INT:Hamiltonian}) in the  $J_K$---$J_H$ space ($t=1$).  For any $J_H>0$, we find a transition from the KI to the algebraic spin liquid.
}
\end{figure}

In this paper we begin by a mean-field study of the Kondo--Heisenberg model Eq.~(\ref{INT:Hamiltonian}) on the honeycomb lattice. We adopt a fermionic representation for localized spins. We make a mean-field decoupling of $\bm{S}_i\cdot\bm{S}_j$ in the fermionic hopping channel. This is sometimes referred to as the resonating valence bond (RVB) decoupling. The Kondo interaction is decoupled in the usual mean-field way with the hybridization ``order parameter" $b=\left< \sum_a f_i^{a \dagger} c_i^a \right>$. The nonzero $b$ will correspond to the KI phase, where the gap in the dispersion is proportional to $b$. 
For any $J_H>0$ we find a continuous transition between the KI ($b\neq0$) and a spin liquid state (see Fig.~\ref{FIG:KI_ASL}).  This spin liquid is characterized by Dirac fermions coupled to U(1) gauge field, and has been called the algebraic spin liquid (ASL), because the spin correlation decays as a power law and spin excitations are gapless.\cite{Rantner02} 

%@@@
As mentioned earlier, in the $J_K \ll J_H$ limit, localized spins will be Neel ordered. Our mean field decoupling can not capture this. \footnote{The issue is that for decoupling the localized spin interactions - in the large N expansion - one either uses the Schwinger boson representation  or the fermion representation. KI is absent in the former formalism and AF ordered phase is absent in the later one.} Here we proceed with the assumption that the phase diagram is given by Fig.~\ref{FIG:KI_SL_Neel} and our main interest is to study the KI to SL transition. For this purpose the RVB decoupling is a reasonable starting point.

%Thus our mean-field theory supports the scenario shown in Fig.~\ref{FIG:KI_SL_Neel}.

The rest of the paper is devoted to studying the critical properties of the fixed point that controls the transition from the KI to ASL. We note that neither phase breaks any physical symmetry, so that the transition is not described by the standard Ginzburg-Landau-Wilson critical point. What changes at the transition is the dynamics of an emergent gauge field. The emergent gauge field is confined in the KI phase and it is deconfined in the ASL phase.

In Sec.~\ref{Large N Formulation} we generalize our model by letting the spin indices to run from 1 to N rather than just ``up'' and ``down'' . The saddle point approximation (i.e. the mean-field) becomes exact as $N \rightarrow \infty$. In Sec.~\ref{SEC:EFT} we derive the low-energy Lagrangian density which sets the stage for a systematic $1/N$ expansion.

Section~\ref{SEC:Feynman_Diagrams} develops the propagator for the Kondo field $b$, and the gauge field $a_{\mu}$ in the leading order. The Kondo field propagator can be tuned to a massless point at $J_K=J_K^{cr}$. In the $J>J_K^{cr}$ regime the $b$ field condenses to $|b|$ and we will be in the Higgs phase where the gauge field is massive. This is nothing but the Kondo physics, where the excitation gap is proportional to $|b|$. In the $J<J_K^{cr}$ regime the Kondo field is not condensed, and is massive. The physics is described by Dirac fermions coupled to a U(1) gauge field. This field theory, known as QED$_{3}$, has been studied extensively\cite{Rantner02, Hermele04, Hermele05, Appel86, Vafek02}, and is understood to be an algebraic spin liquid due to algebraic correlation of spins.\cite{Rantner02} This phase is believed to be deconfined.\cite{Hermele04}

Section~\ref{SEC:nu} contains the main technical calculation of this paper. We calculate the exponent $\nu$, which describes the diverging length scale of the transition. We consider a relativistic fixed point, which allows us to adopt standard field theory methods. In Sec.~\ref{SEC:SSC} we calculate the decay exponent of the staggered localized spin. In Sec.~\ref{SEC:stability} we show that the relativistic fixed point is stable and is therefore the appropriate fixed point to study for the transition.

%*****************************************
\section{Finite Coupling Kondo Transition}
%*****************************************

%*************************************************
\subsection{Mean-Field Transition \label{SEC:MFT}}
%*************************************************
Our starting point is the Hamiltonian
\begin{equation} \label{EQ:Hamiltonian}
\hat{H} = -t \sum_{\left<ij\right>,a} \left( c_{i}^{a\dagger} c_{j}^{a} + H.c. \right) + J_K \sum_i \bm{s}_i\cdot \bm{S}_i + J_H \sum_{\left<ij\right>} \bm{S}_i\cdot \bm{S}_j,
\end{equation}
where $i$ and $j$ live on the honeycomb lattice sites and $a$ is the spin index: $\{\uparrow,\downarrow\}$. The $\bm{s}_i$ denotes the conduction electron spin at the site i and the capital $\bm{S}_i$ denotes the localized spin at the site i and they both are SU(2) spins with an SU(2) spin algebra. The conduction electron spin is given by
\begin{equation}  \label{EQ:s}
\bm{s}_i = \frac{1}{2} c^{a \dagger}_i \bm {\sigma}_{ab} c^b_i,
\end{equation}
where $\bm{\sigma}=\left(\sigma_x,\sigma_y,\sigma_z\right)$ are the Pauli matrices. 
We adopt a fermionic representation for localized spins, where their Hilbert space $\{\ket{\Uparrow_i},\ket{\Downarrow_i}\}$, is constructed using Fermionic operators:
\begin{equation}
\begin{split}
\ket{\Downarrow_i} &= f_i^{\downarrow \dagger}\ket{0}, \\
\ket{\Uparrow_i} &= f_i^{\uparrow \dagger}\ket{0}. 
\end{split}
\end{equation}
The anti--commutation relation:
\begin{equation} \label{EQ:{f,f}}
\{f^{a \dagger}_i,f^b_j\}= \delta_{ab}\delta_{ij} \\
\end{equation}
together with
\begin{equation} \label{EQ:S}
	\bm{S}_i = \frac{1}{2} f^{a \dagger}_i\bm{\sigma}_{ab} f^b_i 
\end{equation}
will result in the SU(2) commutation relations for spin $\bm{S}$. However, the Hilbert space of the \textit{localized} spin $\bm{S}_i$ is restricted to only two elements: $\{\ket{\Uparrow_i},\ket{\Downarrow_i}\}$. Therefore identifying localized spin by fermionic operators given by Eq.~(\ref{EQ:S}) has to be accompanied by the constraint
\begin{equation} \label{EQ:constraint}
\sum_{a}f^{a \dagger}_i f^a_i = 1.
\end{equation}

Combining Eqs.~(\ref{EQ:s}) ---~(\ref{EQ:constraint}); together with the following identity for the Pauli matrices
\begin{equation} 
\bm{\sigma}_{ab} \cdot \bm{\sigma}_{cd} = 2\delta_{ad}\delta_{bc}-\delta_{ab}\delta_{cd}
\end{equation}
one finds a fermionic representation of the spin--spin interactions in the Hamiltonian of Eq.~(\ref{EQ:Hamiltonian}):
\begin{align}	
\label{EQ:s.S} \bm{s}_i \cdot \bm{S}_i  &= - \frac{1}{4}\l[\left(c_{i}^{a \dagger} f_{i}^{a}\right) \left(f_{i}^{b \dagger} c_{i}^{b}\right)+ H.c.\r] + \frac{1}{4}, \\ 
\label{EQ:S.S} 	\bm{S}_i \cdot \bm{S}_j &= - \frac{1}{4}\l[\left(f_{i}^{a \dagger} f_{j}^{a}\right) \left(f_{j}^{b \dagger} f_{i}^{b}\right)+H.c. \r] + \frac{1}{4},
\end{align}
where the sum over the spin indices is understood.

The four-fermion interaction terms of Eqs.~\ref{EQ:s.S} and~\ref{EQ:S.S} will be replaced in the mean-field (MF) simplification by a  quadratic interaction:
\begin{equation} \label{EQ:MFdecomposition} 
\hat{A} = \hat{A_1}\hat{A_2} \rightarrow \left<A_1\right> \hat{A_2}+ \hat{A_1} \left<A_2\right>	- \left<A_1\right> \left<A_2\right>,
\end{equation}
where $\hat{A}$ denotes a generic four-fermion interaction term. To make the interaction quadratic one has to ignore $(A_1-\left<A_1\right>)(A_2-\left<A_2\right>)$. So the mean-field parameters have to satisfy the  self-consistency condition:
\begin{equation}\label{EQ:SC_condition}
\frac{\partial F_{MF}}{\partial \left<A_i\right>}=\left<\hat{A}_i-\left<A_i\right>\right>=0.
\end{equation}

We make the interaction terms given by Eqs.~(\ref{EQ:s.S}) and~(\ref{EQ:S.S}) quadratic by choosing the following mean-field parameters:
\begin{align} 	
\label{EQ:MF:b}   \left<\sum_a c_{i}^{a \dagger} f_{i}^{a}\right> &= -|b_i|e^{i\theta_i},  \\ 
\label{EQ:MF:chi} \left<\sum_a f_{i}^{a \dagger} f_{j}^{a}\right> &= -|\chi_{ij}| e^{i a_{ij}}.
\end{align}
The first parameter $|b_i|e^{i\theta_i}$ lives on sites.  The second parameter $|\chi_{ij}| e^{i a_{ij}}$ lives on links.  
We assume  that these parameters have a constant magnitude throughout the lattice. When they take a non-zero value, small deviations from their ``constant" magnitudes cost energy: they are locally stable. $|b_i|=b\neq 0$ corresponds to the Kondo insulator phase, where the gap in the excitations is proportional to $b$.  We also note that the conventional decoupling of $\bm{S}_i \cdot \bm{S}_j$ leads to the anti-ferromagnetic order parameter $\left<S_{iz}\right>$. In this paper we adopt the alternative decoupling of Eq.~(\ref{EQ:MF:chi}), which is often called the RVB (Resonating Valence Bond) decoupling. RVB decoupling leads to a spin liquid state.

The U(1) gauge freedom in defining $f$ gives the freedom to eliminate $\theta_i$ by the following gauge transformation:
\begin{equation} \label{EQ:gauge}
\begin{split}
f_i^a &\rightarrow e^{i\theta_i} f_i^a ,  \\
a_{ij} &\rightarrow a_{ij} + \left(\theta_i-\theta_j\right).
\end{split}
\end{equation}
This transformation leaves the total flux modulo $2\pi$
\begin{equation}
\sum_{\hexagon}a_{ij}\ (\rm{mod}\  2\pi)
\end{equation} through any closed loop invariant. Next we make a choice $\chi_{ij}=\chi$ where $\chi$ is real. This choice corresponds to the case where the total flux through each hexagon modulo $2\pi$ is zero. 

Our mean-field choices result in the following mean-field Hamiltonian:
\begin{equation} \label{EQ:MFHamiltonian}
\begin{split}
\hat{H}_{MF}&= -t \sum \left( c_{i}^{a\dagger} c_{j}^{a} + H.c. \right)  \\
		  &\ +\frac{J_K b}{2}\sum \left(f_{i}^{a\dagger} c_{i}^{a}+H.c.\right)  \\
		  &\ + \frac{J_H\chi}{2} \sum \left(f_{i}^{a\dagger} f_{j}^{a}+ H.c.\right)  \\
		  &\ + \mathcal{N}\left(J_K b^2 +\frac{3J_H}{2} \chi^2\right) ,
\end{split}
\end{equation}
where $\mathcal{N}$ is the number of unit cells. The different coefficients of the constant terms are due to the fact that there are 2 sites as opposed to 3 links per unit cell. We also mention that, at the mean- field level, the constraint given by Eq.~(\ref{EQ:constraint}) is enforced on average by having a chemical potential for $f$ fermions. This chemical potential is zero due to f-fermions particle-hole symmetry.

From $\hat{H}_{MF}$ , the MF free energy can be calculated numerically. MF parameters satisfying the self consistency condition are then obtained. The numerics confirm the finite coupling Kondo transition in a background $\chi$, for any $J_H>0$. Next we show this finite coupling Kondo transition analytically and find the exponent $\beta$
\begin{equation}
|b| \propto (J_K-J_K^{cr})^{\beta},
\end{equation}
by focusing on the low-energy theory near the Dirac nodes.

We focus first on the $c$ electrons kinetic term  $-t \sum \left( c_{i}^{a\dagger} c_{j}^{a} + H.c. \right)$ . We do the Fourier transformation
\begin{equation}
c_A(\bm{k})=\frac{1}{\mathcal{N}}\sum_{\bm{R}} e^{i\bm{k}\cdot \bm{R}} c_A(\bm{R})
\end{equation}
where $\bm{R}$ are the Bravais lattice vectors and $c_A(\bm{R})$ is the conduction electron annihilation operator at the $A$ sublattice site of the unit cell positioned at $\bm{R}$. The similar transformation is done to get $c_B(\bm{k})$. Conduction electrons kinetic term in momentum space will take the form:
\begin{equation} \label{EQ:c_kinetic_begin}
-\sum_{\left<ij\right>} t c_i\d c_j + H.c. = - t \sum_{\bm k} \Gamma(\bm{k}) c_A\d(\bm{k}) c_B(\bm{k}) + H.c. ,
\end{equation}
where
\begin{equation}
\Gamma(\bm{k}) =   1 + e^{i\bm{k}\cdot (\bm{\eta}_2-\bm{\eta}_1)} +e^{-i\bm{k}\cdot (2\bm{\eta}_1+\bm{\eta}_2)},
\end{equation}					
and $\bm{\eta}_1=l (1,0)$ and $\bm{\eta}_2=l (-1/2, \sqrt{3}/2)$ are two of the three vectors that connect $A$ sublattice sites to nearest neighbor $B$ sublattice sites. $l$  is the length between nearest neighbor sites which will be set to the unit length: $l=1$.

$\Gamma(\bm{k})$ vanishes at the Dirac nodes $\pm \bm{k}_D$, which is given by
\begin{equation}
\bm{k}_D=\bigl(0,\frac{4\pi}{3\sqrt{3}}\bigr).
\end{equation}

\begin{figure}
\begin{center}
\includegraphics{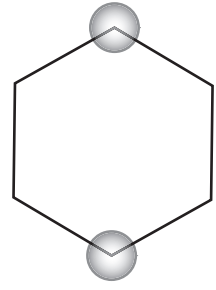}
\end{center}
\caption{ \label{FIG:Brillouin-Zone}
The Brillouin zone of the honeycomb lattice. Corners of  the Brillouin zone are where the tight-binding gap vanishes. The \textit{independent } low-energy modes are denoted by the filled-circles around the two Dirac nodes $\pm \bm{k}_D$.
}
\end{figure}

Near the nodes as schematically represented by the filled circles in the Fig.~\ref{FIG:Brillouin-Zone}, $\Gamma(\pm \bm{k}_D+\bm{q})$ is given by
\begin{equation} \label{EQ:gamma(kd+l)}
\Gamma(\pm \bm{k}_D +\bm{q})=\frac{3}{2}(i q_1 \mp q_2)+ \mathcal{O}(\bm{q}^2).
\end{equation}
To write the conduction electron kinetic term near the Dirac nodes, we use the following notation, which will also be helpful  in Sec.~\ref{SEC:EFT}:
\begin{equation} \label{EQ:def_1}
\psi_{\pm}(\bm{q}) \equiv
\begin{bmatrix} 
  c_A(\pm\bm{k}_D+\bm{q})\\ c_B(\pm\bm{k}_D+\bm{q}).
\end{bmatrix}	
\end{equation}
After collecting Eqs.~(\ref{EQ:c_kinetic_begin}) ---~(\ref{EQ:def_1}), we get
\begin{equation}
\begin{split}
-\sum_{\left<ij\right>} t c_i\d c_j + H.c. &\sim  \frac{3 t}{2} \sum_{\bm{q}} \psi_+(\bm{q})\d(q_1\sigma_2+q_2 \sigma_1) \psi_+(\bm{q}) \\
&+ \frac{3 t}{2} \sum_{\bm{q}} \psi_-(\bm{q})\d(q_1\sigma_2-q_2 \sigma_1) \psi_-(\bm{q}),
\end{split}
\end{equation}
where ``$\sim$'' is used, since in contrast to Eq.~(\ref{EQ:c_kinetic_begin}) the momentum sum is restricted to be near the Dirac nodes. 

Next we obtain the energy dispersion near the node $+\bm{k}_D$.  The MF Hamiltonian in momentum space and near the node $+\bm{k}_D$, denoted by $\hat{H}_{+}(\bm{q})$, is given by 
\begin{equation}
\hat{H}_{+}(\bm{q})=
\begin{bmatrix} \psi_{+}(\bm{q})\d & \phi_{+}(\bm{q})\d \end{bmatrix}
\mathcal{H}_+
\begin{bmatrix} \psi_{+}(\bm{q})\\ \phi_{+}(\bm{q})\\ \end{bmatrix},
\end{equation}
where we have dropped the constant term of Eq.~(\ref{EQ:MFHamiltonian}) and we have used a similar notation as in Eq~(\ref{EQ:def_1}) for the $f$ fermions:
\begin{equation}\label{EQ:def_2}
\phi_{\pm}(\bm{q}) \equiv \begin{bmatrix} f_A(\pm\bm{k}_D+\bm{q})\\ f_B(\pm\bm{k}_D+\bm{q})\\ \end{bmatrix}.
\end{equation}
The $4\times4$ matrix $\mathcal{H}_+$ is given in block form by 
\begin{equation}
\mathcal{H}_+ = \frac{1}{2}
\begin{bmatrix} 
2v_c(q_1 \sigma_2 + q_2 \sigma_1) & J_K b \\%\ \mathbf{1}_{2\times2}\\
J_K b  & -2v_f(q_1 \sigma_2 + q_2 \sigma_1)	\\					
\end{bmatrix},
\end{equation}
where $v_c$ and $v_f$ are given by:
\begin{eqnarray}
v_c &=& \frac{3}{2} t , \\
v_f &=& \frac{3}{4}\chi J_H.
\end{eqnarray}
It will be clear why notations $v_c$ and $v_f$ is used shortly.
It is straightforward to diagonalize the above matrix. The four eigenvalues are given by:
\begin{equation}
e(\bm{q})= \pm \left((v_c-v_f)|\bm{q}| \pm \sqrt{(v_f+v_c)^2 \bm{q}^2 + J_K^2 b^2}\right)/2.
\end{equation}
We have dropped the subscript $+$ from $e(\bm{q})$, since the dispersion near the other node is the same. 

To understand the above dispersion better, we mention its behavior in different regimes:
\begin{itemize}
\item $b=0$. The MF Hamiltonian has two decoupled parts: kinetic terms for the conduction electrons and for f fermions. In this case the 4 eigenvalues near the nodes become $\pm v_c |q|$ and $\pm v_f |q|$. They are the massless Dirac dispersions for conduction electrons $c$ and fermions $f$ with \textit{velocities} $v_c$ and $v_f$ respectively. 
The phase with $b=0$ is an algebraic spin liquid phase as will be clear in Sec.~\ref{SEC:SSC}. This is closely related to the work of Rantner and Wen, where Dirac fermions coupled to a gauge field was found to be an algebraic spin liquid.\cite{Rantner02}
\item $v_f=v_c, b\neq 0$. The dispersion is the \textit{massive relativistic} Dirac dispersion 
\begin{equation} \label{EQ:relativistic_dispersion} e(\bm{q})=\pm \sqrt{v_c^2 \bm{q}^2 + (J_K b /2)^2}, \end{equation}
where each eigenvalue is doubly degenerate. $J_K b /2$ is the mass of our relativistic quasi-particles. In Sec.~\ref{SEC:EFT}, we will see that the low-energy theory, in the limit $v_f=v_c$, is Lorentz invaraint. The above relativistic dispersion is a sign of this Lorentz	 invariance. 
\item \textit{sign of} $v_f$ . The sign of $\chi$ ($v_f$) has dramatic consequences on the shape of the dispersion and also on the continuum formulation of the model near the phase transition. Due to the coupling between f-fermions and the conduction electrons the free energy is not invariant under the mapping $\chi \mapsto - \chi$. In the Kondo insulating phase ($b \neq 0$); the dispersion for $v_f>0$ is gapped. However the dispersion for $v_f<0$ has a gapless ring around each Dirac node with the radius
\begin{equation}  q_c=\frac{J_K b }{2\sqrt{-v_c v_f}}. \end{equation} It turns out that the positive (self consistent) $\chi$ ($v_f>0$) is \textit{more stable}. This is not surprising from the above fact that the dispersion for $v_f>0$ is gapped as opposed to $v_f<0$ which is gapless. We are only concerned with the stable solution $\chi>0$ in this paper.
\end{itemize}

The MF transition at zero temperature can be analyzed by
minimizing the total energy
\begin{equation} E=4\sum_{e(\bm{q})<0} e(\bm{q}) + \mathcal{N}\left(J_K b^2 +\frac{3J_H}{2} \chi^2\right) \label{EQ:mf_transition_2}, \end{equation}%
\begin{equation} \frac{\partial E}{\partial b} =0,\ \rm{and}\ \ \frac{\partial E}{\partial \chi} =0 \label{EQ:mf_transition_1}.\end{equation}%
The coeffiecient $4=2\times2$ is for counting spins and the two nodes. The condition $\frac{\partial E}{\partial \chi} =0 $ at $b=0$  results in the self consistent $\chi$, which varies slowly across the transition. For $v_f>0$, the negative energy bands are given by $\left( \pm (v_c-v_f)|\bm{q}| - \sqrt{(v_f+v_c)^2 \bm{q}^2 + J_K^2 b^2}\right)/2$. Therefore 
\begin{equation}  \label{EQ:mf_transition_3}
\sum_{e(\bm{q})<0}e(\bm{q}) = -\sum_{\bm{q}} \sqrt{(v_f+v_c)^2 \bm{q}^2 + J_K^2 b^2}.
\end{equation}
Combine Eqs.~\ref{EQ:mf_transition_2} ---~\ref{EQ:mf_transition_3}, to arrive at the self-consistency equation for $b$:
\begin{equation}
\frac{2}{\mathcal{N}}\sum_{\bm{q}} \frac{J_K}{ \sqrt{(v_f+v_c)^2 \bm{q}^2 + J_K^2 b^2}} = 1 .
\end{equation}
It is clear from the above equation that the Kondo phase does not survive down to $J_K=0$, in contrast to the case when the conduction band possess a Fermi \textit{surface}. The Kondo insulator phase disappears at a \textit{finite} coupling constant $J_K^{cr}$. Taking the limit $\mathcal{N} \rightarrow \infty$, we find the following dependence of the Kondo parameter $|b|$ as a function of the distance from the transition $J_K-J_K^{cr}$: 
\begin{equation}
\begin{split}
|b| &\propto (J_K-J_K^{cr}),  \\
J_K^{cr} &= \frac{v_c+v_f}{\Lambda},
\end{split}
\end{equation}
where $\Lambda \ll 1$ is a large-momentum cutoff around the Dirac nodes $\pm \bm{k}_D$.

To summarize, the MF solution has two interesting features:
\begin{itemize}
\item The Kondo insulator phase disappears at a finite coupling constant.\ 
\item The Kondo condensate $|b|$ vanishes linearly as the distance from the critical point.\ 
We emphasize that a Landau--type expansion for the $b$ mean-field parameter %in a gaussian expansion %valid for a gaussian fixed point 
would yield the result $|b| \propto (J_K-J_K^{cr})^{1/2}$ near the transition point.\ 
The appearance of nontrivial critical exponent at the mean-field level is characteristic of the coupling of $b$ to gapless fermions which is addressed systematically in the $1/N$ expansion.
\end{itemize}

%***********************************************************************************
\subsection{ Large N Formulation : Mean-Field Justified \label{Large N Formulation}}
%***********************************************************************************
In this section we extend the physical model with two spin-flavors to a generalized model with N spin-flavors.\cite{Marston89}\ 
We show that the mean-field solution is the stable solution in the infinite N limit.\ 

The conduction electrons kinetic term $\sum \left( c_i^{a \dagger} c_j^a + H.c. \right)$ is already in the generalized form. We just let the spin index $a$ to run from 1 to N. The spin--spin interactions as written in Eqs.~\ref{EQ:s.S} and~\ref{EQ:S.S} is also easily generalized by letting the spin indices to run from 1 to N. In addition we \textit{control} these interactions by replacing $\frac{1}{2}$ with $\frac{1}{N}$. So the generalized version of the Hamiltonian of Eq.~(\ref{EQ:Hamiltonian}) is given by
\begin{equation}
\begin{split}
\hat{H} &= - t \sum_{\left<ij\right>} \left( c_i^{a \dagger} c_j^a + H.c. \right) \\
&- \frac{J_K}{2N}\sum_{i}\l[(c_{i}^{a\dagger} f_{i}^a)(f_{i}^{b \dagger} c_{i}^b) + H.c.\r]\\
&- \frac{J_H}{2N}\sum_{\left<ij\right>}\l[(f_{i}^{a \dagger} f_{j}^a)(f_{j}^{b \dagger} f_{i}^b)+ H.c.\r],
\label{EQ:generalized_hamiltonian}
\end{split}
\end{equation}
where the sum over spin indices is understood. The above Hamiltonian has to accompanied by the local constraint:
\begin{equation} \label{EQ:large_n_constraint}
\sum_{a}f_{i}^{a \dagger} f_{i}^a = \frac{N}{2}.
\end{equation}
The above generalized model will coincide with the Hamiltonian Eq.~(\ref{EQ:Hamiltonian}) for N=2. In fermion coherent state representation, using Grassmann variables, the Euclidean Lagrangian and the partition function take the form
\begin{equation}
\begin{split}
L_E([c,f,a_0])&= \sum_{i} \Biggl( \bar{c}_i^a(x_0) \partial_{0} c_i^a(x_0) + \bar{f}_i^a(x_0) \partial_{0} f_i^a(x_0) \\
&\  -i a_{i0} \left(\bar{f}_i^a(x_0) f_i^a(x_0)-\frac{N}{2}\right) \Biggr)+ H,  \\
Z &= \int \mathcal{D}[\bar{c}c\bar{f}f a_0] \ e^{-\int dx_0 L_E(x_0)},
\end{split}
\end{equation}
where $x_0$ denotes the imaginary time. The field $a_{i0}$, when integrated over, will produce delta functions at each site enforcing the local constraint given by Eq.~\ref{EQ:large_n_constraint}. The notation $a_{i0}$ is used because it will serve as the \textit{time} component of the emergent gauge field.

Next we do the standard Hubbard--Stratonovich transformation to decouple the four-fermion terms in the action at a cost of introducing the complex fields $b_i$, which live on sites and $\chi_{ij}$, which live on links:
\begin{equation}
\begin{split}
&L_E([c,f,b,\chi,a_0])   \\
&=\sum_i  \bar{c}_i^a \partial_{0} c_i^a -t \sum_{\left<ij\right>}\left(\bar{c}_i^a c_j^a+ c.c.\right) \\
&+ \sum_i \Bigl(
\bar{f}_i^a \left(\partial_{0}-ia_{i0}\right) f_i^a  + iNa_{i0} /2 \Bigr)  \\ % 
&+  \sum_{\left<ij\right>} \Bigl(\frac{N}{J_H}|\chi_{ij}|^2+ \chi_{ij} \bar{f}_i^a f_j^a + \chi_{ij}^* \bar{f}_j^a f_i^a \Bigr) \\
&+ \sum_i \Bigl( \frac{N}{J_K} |b_i|^2 + b_i \bar{c}_i^a f_i^a  + b_i^* \bar{f}_i^a c_i^a\Bigr), \\
\\
Z&=\int \mathcal{D}[\bar{c} c \bar{f} f b^* b \chi^* \chi a_0]e^{-\int dx_0 L_E(x_0)}. \\
\label{EQ:microscopic lagrangian}
\end{split}
\end{equation}
The $x_0$ dependence of Grassmann and complex fields in the Lagrangian is understood as well as the dropped indices in the measure of the path integral.

The action is quadratic in the Fermionic fields and it can be integrated out at the cost of obtaining an effective action for $b_i$ and $\chi_{ij}$ fields, which is highly \textit{nonlocal}. Since there is no mixing of different spin-flavors, the resulting effective action will be proportional to $N$:
\begin{equation}
%L_E([b,\chi]) = N \left( \log \det \mathcal{M} + \frac{1}{J_K}\sum_i |b_i|^2 + \frac{1}{J_H}\sum_{\left<ij\right>} |\chi_{ij}|^2 \right).
\frac{L_E([b,\chi])}{N} = \log \det \mathcal{M} + \frac{1}{J_K}\sum_i |b_i|^2 + \frac{1}{J_H}\sum_{\left<ij\right>} |\chi_{ij}|^2 .
\end{equation}
$\mathcal{M}$ is a $4\mathcal{N} \times 4\mathcal{N}$ Hermitian matrix, which encodes the quadratic interaction of the fermionic fields as given by Eq.~(\ref{EQ:microscopic lagrangian}), e.g. $\mathcal{M}_{\bar{c}_i f_i}=b_i$.\cite{Negele88} The ``$\det \mathcal{M}$" makes the effective Lagrangian in terms of $b_i$ and $\chi_{ij}$ fields nonlocal.

The saddle point approximation for the above Lagrangian $L_E([b,\chi])$ is exact as $N\rightarrow \infty$. The saddle point equations will be the mean-field equations given by Eq.~(\ref{EQ:SC_condition}), and this is how the mean-field developed in the previous section is justified. Furthermore the Lagrangian given in Eq.~(\ref{EQ:microscopic lagrangian}) will set the stage for a systematic $1/N$ expansion around the mean-field solution. 

%************************************************************
\section{Beyond Mean-field: low-energy Effective Field Theory \label{SEC:EFT}}
%************************************************************

In this section we obtain the Lagrangian density near the critical point starting from the microscopic Lagrangian. There are two independent nodes  $\pm \bm{k}_D$ and near each node there are fermions from $A$ and $B$ sublattices. We package these $c$ and $f$ fermions into four component spinors $\Psi$ and $\Phi$. This is done in a way to make the kinetic terms for both fields look the same. %Otherwise we have to use different notations for $\Psi$ and $\Phi$ propagators. We will make an assumption about the strength of $J_H$ to make the final field theory Lorentz invariant. 

Since there is no mixing of different spin-flavors in the microscopic Lagrangian Eq.~(\ref{EQ:microscopic lagrangian}), we will drop this index throughout the derivation of the Lagrangian density. The spin-flavor index will be resurrected at the end.

The kinetic term for the conduction electrons was discussed in Sec.~\ref{SEC:MFT}:
\begin{equation}  \label{EQ:c_hopping}
\begin{split}
- t \sum  c_i\d c_j + H.c. &\sim  v_c \sum_{\bm{q}}\psi_+(\bm{q})\d(q_1\sigma_2+q_2 \sigma_1) \psi_+(\bm{q}) \\
&+ v_c \sum_{\bm{q}} \psi_-(\bm{q})\d(q_1\sigma_2-q_2 \sigma_1) \psi_-(\bm{q})
\end{split}
\end{equation}

Note that under the transformation $\psi_{-}(\bm{q}) \rightarrow \sigma_2 \psi_{-}(\bm{q})$ the second term has the same form as the first one. We combine the \textit{two} 2-component spinors, associated with the two Dirac nodes, to form a 4-component spinor $\Psi(\bm{q})$:
\begin{equation}
\Psi(\bm{q}) \equiv
\begin{bmatrix}
\psi_+(\bm{q}) \\
\sigma_2 \psi_-(\bm{q}) \\
\end{bmatrix}.
\end{equation}
To write the kinetic term in terms of $\Psi(\bm{q})$, we define the $4\times4$  \textit{block diagonal} matrices $\tau_i$:
%\begin{bmatrix}
% c_{A +}(\bm{q})\\
% c_{B+}(\bm{q})\\
% -i c_{B-}(\bm{q}) \\
% i c_{A-}(\bm{q})\\
%\end{bmatrix} \nonumber \\
\begin{equation}
\tau_i  \equiv  
\begin{bmatrix}
\sigma_i & 0 \\
0 & \sigma_i \\
\end{bmatrix} = 
\sigma_i \otimes \mathbf{1}_{2\times2}.
\end{equation}
The kinetic term then takes the form
\begin{equation} 	\label{EQ:kinetic_c_Psi}
-t \sum c_i\d c_j + H.c.  \sim  v_c \sum_{q} \Psi(\bm{q})\d (q_1 \tau_2 + q_2 \tau_1) \Psi(\bm{q}).  \\
\end{equation}

After taking the continuum limit, the Lagrangian density corresponding to this Hamiltonian is:
\begin{equation}
\mathcal{L}_E^{(1)}=\Psi\d \partial_0 \Psi  - i v_c \Psi\d  (\tau_2\partial_1+\tau_1\partial_2)\Psi.
\end{equation}
To use the field theory techniques one would write $\mathcal{L}_E^{(1)}$ in a different form
\begin{equation} 	\label{EQ:L_E1}
\mathcal{L}_E^{(1)}= \bar{\Psi} \tau_3 \partial_0 \Psi  + v_c \bar{\Psi}  (-\tau_1\partial_1+\tau_2\partial_2)\Psi,
\end{equation}
where  $\bar{\Psi}$ is defined to be
\begin{equation} \label{DEF:psi_bar}
 \bar{\Psi} \equiv \Psi\d \tau_3 .
\end{equation}

Next we focus on $\sum \left( \chi_{ij} \bar{f}_i^a f_j^a + \chi_{ij}^{*} \bar{f}_j^a f_i^a \right)$ term in Eq.~(\ref{EQ:microscopic lagrangian}). Consider the mean-field case $\chi_{ij}=\chi>0$. The phase fluctuations will be included as gauge field fluctuations later on. Ignoring the phase fluctuations, this term looks exactly like the kinetic term  of the conduction electrons given by Eq.~(\ref{EQ:c_hopping}), but with an opposite sign:
\begin{equation}
\begin{split}
\chi \sum  f_i\d f_j + H.c. &\sim  v_f \sum_{\bm{q}} \phi_+ (\bm{q}) \d(-q_1\sigma_2-q_2 \sigma_1) \phi_+ (\bm{q})\\
&+ v_f \sum_{\bm{q}} \phi_-  (\bm{q}) \d(-q_1\sigma_2+q_2 \sigma_1) \phi_- (\bm{q}). 
\end{split}
\end{equation}
We make this look like Eq.~(\ref{EQ:kinetic_c_Psi}) by a further transformation to absorb the minus sign and we again combine the 2-component spinors near the two nodes to form a 4-component spinor $\Phi(\bm{q})$: 
\begin{equation}
\Phi(\bm{q}) \equiv
\begin{bmatrix}%
\sigma_3 \phi_+ (\bm{q})\\
-i\sigma_1 \phi_- (\bm{q})\\
\end{bmatrix}.
\end{equation}
On the grounds of gauge invariance, we write the Lagrangian density corresponding to the ``kinetic" term of the $f$ fermions
\begin{equation} 	\label{EQ:L_E2}
\mathcal{L}_E^{(2)} = \bar{\Phi}\tau_3 (\partial_0-ia_0) \Phi + v_f \bar{\Phi} \Bigl(-\tau_1(\partial_1-ia_1)+\tau_2(\partial_2-ia_2)\Bigr)\Phi,
\end{equation}
where the non-compact gauge field $\bm{a}(\bm{x})$ appears as a connection to produce covaraint derivative. It is related to the the \textit{phase} $a_{ij}$ through the relation $\bm{a}(\bm{x})=a_{ij}\hat{\eta}_{ij}$, where $\hat{\eta}_{ij}$ is the unit vector connecting the site $i$ to the site $j$. We have used the same definition as in Eq.~(\ref{DEF:psi_bar}):
\begin{equation} \label{DEF:phi_bar}
\bar{\Phi} \equiv \Phi\d \tau_3 .
\end{equation}
En route to the continuum limit, we have implicitly dropped the compact character of the gauge field. This is only valid if the gauge fluctuations are not very strong. In the imaginary time evolution, there will be events on the lattice scale that change the flux of the gauge field by $2\pi$. If these events do not proliferate, we will be in small gauge coupling constant regime and taking the continuum limit is justified.\cite{Kogut79} We return to this point at the end of this section.

Now we focus on the interaction term $\sum_i b_i c_i \d f_i$. We write this in the momentum space and restrict $c$ and $f$ fermions momentum to be near $\pm \bm{k}_D$, and $b$ momentum to be near $k=0$. Simple manipulations
\begin{eqnarray}
\psi_+\d(\bm{q})  \phi_+(\bm{q}') &=& \psi_+\d(\bm{q}) \sigma_3 \left(\sigma_3 \phi_+(\bm{q}')\right), \\
\psi_- \d(\bm{q}) \phi_-(\bm{q}') &=& \left(\sigma_2 \psi_-(\bm{q})\right)\d \sigma_3 \left(-i\sigma_1 \phi_-(\bm{q}')\right),
\end{eqnarray}
combined with the definitions for $\Psi$ and $\Phi$ fields, result in
\begin{equation}
\begin{split}
\psi_+\d(\bm{q})  \phi_+(\bm{q}') + \psi_- \d(\bm{q}) \phi_-(\bm{q}') &= \Psi\d(\bm{q}) \tau_3 \Phi(\bm{q}') \\
&= \bar{\Psi}(\bm{q})\Phi(\bm{q}')  .
\end{split}
\end{equation}
Therefore in the continuum limit the Lagrangian density corresponding to $\sum_i \left(b_i c_i \d f_i+b_i^{*}f_i \d c_i\right)$ will be
\begin{equation} \label{EQ:L_E3}
\mathcal{L}_E^{(3)}=b\bar{\Psi} \Phi + b^*\bar{\Phi} \Psi.
\end{equation}

Now collect $\mathcal{L}_E^{(1)}$, $\mathcal{L}_E^{(2)}$ and $\mathcal{L}_E^{(3)}$, to write the Lagrangian density %corresponding to the microscopic Lagrangian of Eq.~(\ref{EQ:microscopic lagrangian}):
\begin{equation} \label{EQ:L_E_no_Lorentz}
\begin{split}
\mathcal{L}_E^{v_c/v_f} &= \bar{\Psi}^a \gamma_{\mu}^{v_c/v_f} \partial_{\mu} \Psi^a + \bar{\Phi}^a \gamma_{\mu}\left(\partial_{\mu}-ia_{\mu}\right) \Phi^a  \\
&+ b\bar{\Psi}^a \Phi^a + b^*\bar{\Phi}^a \Psi^a + \frac{N}{J_K} |b|^2 + \cdots ,
\end{split}
\end{equation}
where the spin indices are restored. $\gamma_{\mu}^\varrho$ is defined to be
\begin{equation} 	\label{EQ:gamma_mu(v)}
\gamma_{\mu}^\varrho \equiv \left(\tau_3,- \varrho \tau_1, \varrho\tau_2\right),
\end{equation}
and $\gamma_{\mu}$ is a shorthand for $\gamma_{\mu}^\varrho$ with $\varrho=1$.\ 
The velocity $v_f$ in Eq.~(\ref{EQ:L_E2}) is absorbed by scaling $\bm{x}$ and $\bm{a}$. As a result, the velocities enter only as the dimensionless ratio $v_c/v_f$ via $\gamma_{\mu}^{v_c/v_f}$.

Taking the continuum limit, which is essentially a coarse graining process, generates new interactions among coarse grained fields.\cite{Ma76} The ellipsis in Eq.~(\ref{EQ:L_E_no_Lorentz}) denotes such new interactions. Not all possible terms are allowed though. Since the coarse graining process does not break symmetries of the microscopic Lagrangian, the generated interactions should respect these symmetries.\cite{Wen02}

Next we set $v_f=v_c$ to make the low-energy theory Lorentz symmetric. The action in the presence of \textit{two different velocities} will not respect the Lorentz symmetry. Under RG transformation --- in the absence of Lorentz symmetry --- the velocities \textit{will} change, since there is no symmetry to protect them from changing. We shall show later that the Lorentz invaraince is protected in the RG sense, in that small deviation from $v_f=v_c$ will scale to zero under RG transformation. Thus we focus our discussion on this Lorentz invariant fixed point.

%We emphasize that to obtain the critical properties of a second order phase transition one has to be close to a stable fixed point. One might obtain logarithmic singularities for some propagators starting from the Lagrangian density given by Eq.~(\ref{EQ:L_E_no_Lorentz}). But it would be wrong to translate these logarithmic singularities to critical exponents, since  the action for $v_f\neq v_c$ is not close to the fixed point action. 

The Lorentz-symmetric Lagrangian density is given by:
\begin{equation} \label{EQ:L_E}
\begin{split} 
\mathcal{L}_E &= \bar{\Psi}^a \gamma_{\mu} \partial_{\mu} \Psi^a + \bar{\Phi}^a \gamma_{\mu}\left(\partial_{\mu}-ia_{\mu}\right) \Phi^a \\
&+b\bar{\Psi}^a \Phi^a + b^*\bar{\Phi}^a \Psi^a + \frac{N}{J_K} |b|^2  \\
&+\frac{N}{2e^2}\left( \epsilon_{\mu\nu\lambda}\partial_{\nu}a_{\lambda} \right)^2+N g |\left(\partial_{\mu}+ia_{\mu}\right)b|^2+\cdots, 
\end{split}
\end{equation}
where $\mathcal{L}_E$ is a shorthand for $\mathcal{L}_E^{\varrho}$ with $\varrho=1$. The last two terms are two examples of the terms that will be generated in the coarse graining process. They are written on the grounds of gauge invariance. In Sec.~\ref{SEC:stability} we show that these new terms and other terms grouped in the ellipsis are irrelevant at our finite coupling Kondo transition fixed point. Embedded in Eq.~(\ref{EQ:L_E}), is the QED$_{3}$ Lagrangian desnity
\begin{equation} 	
\bar{\Phi}^a \gamma_{\mu}\left(\partial_{\mu}-ia_{\mu}\right) \Phi^a
+\frac{N}{2e^2}\left( \epsilon_{\mu\nu\lambda}\partial_{\nu}a_{\lambda} \right)^2 + \cdots  \ . 
\end{equation}
QED$_{3}$ has been studied before as the low-energy theory of the staggered flux phase \cite{Rantner02, Hermele04, Hermele05}. We have used the recent result\cite{Hermele04} about the irrelavance of instanton events near the QED$_{3}$ fixed point and took the gauge field to be non-compact. In our formulation the \textit{physical} SU(2) theory corresponds to N=2 \textit{four}-component spinors. It could have been formulated as $N_f=4$ \textit{two}-component spinors near each node in the Brillouin zone. To go back and forth between these formulations one just needs to use $N_f=2 N$.
%\newpage

\begin{fmffile}{fmfkhh}

%******************************************************
\section{$1/N$ Systematic Expansion: Field Propagators \label{SEC:Feynman_Diagrams}}
%******************************************************

The Lagrangian density Eq.~(\ref{EQ:L_E}), derived in the previous section, allows us to develop the machinery of $1/N$ expansion. The propagators for the fields $b$ and $a_{\mu}$ are of order $1/N$ in the large $N$ limit. This will alow us to calculate different propagators and scaling dimensions order by order in $1/N$.

We start with our notations for propagators of the fermion fields, $\Psi$ and $\Phi$. We use a single line notation for the $\Phi$ field propagator and a double--line for $\Psi$ propagator:
\begin{eqnarray}
\left<\Psi_{\rm{i}}^a \bar{\Psi}_{\rm{j}}^b\right>(q) &=&
{\rm i}\ \parbox{20mm}{\begin{fmfgraph*}(20,12)
\fmfstraight
\fmfleft{i}
\fmfright{o}
\fmf{dbl_plain_arrow,label=$q$}{o,i}
\end{fmfgraph*}}\ {\rm j}
= \delta^{ab}\left(i\slashed{q}\right)^{-1}_{\rm{ij}}, \\
\left<\Phi_{\rm{i}}^a \bar{\Phi}_{\rm{j}}^b\right>(q) &=&
{\rm i}\ \parbox{20mm}{\begin{fmfgraph*}(20,12)
\fmfstraight
\fmfleft{i}
\fmfright{o}
\fmf{plain_arrow,label=$q$}{o,i}
\end{fmfgraph*}}\ {\rm j}
= \delta^{ab}\left(i\slashed{q}\right)^{-1}_{\rm{ij}}, \\
\nonumber
\end{eqnarray}
where Feynman's slash notation
\begin{equation}
\slashed{q}=q_{\mu}\gamma_{\mu}
\end{equation}
is used.\\

\textit{The Gauge Field $a_{\mu}$}. Our Lagrangian density $\mathcal{L}_E$ has a QED$_{3}$ piece 
\begin{equation} 	\label{EQ:qed_3}	
\bar{\Phi}^a \gamma_{\mu}\left(\partial_{\mu}-ia_{\mu}\right) \Phi^a 	+\frac{N}{2e^2}\left( \epsilon_{\mu\nu\lambda}\partial_{\nu}a_{\lambda} \right)^2 + \cdots,
\end{equation}
which has been studied before.\cite{Hermele04, Hermele05, Rantner02, Appel86, Vafek02} We refer the reader to the Appendix B of Ref.~\onlinecite{Hermele05} for the calculation of the  QED$_{3}$  gauge propagator. The QED$_{3}$ gauge propagator, in the leading order, is given by:%\footnote{Note that the different factors (8 v.s. 16) is because of $N_f$ in Hermele's paper is equal to $2 N$.}:
\begin{equation}
\left<a_{\mu}a_{\nu}\right>(q)=\mu\ 
 \parbox{20mm}{\begin{fmfgraph*}(20,12)
\fmfstraight
\fmfleft{i}
\fmfright{o}
\fmf{photon,label=$q$}{o,i}
\end{fmfgraph*}}\ \nu =\frac{8}{N|q|}\biggl(\delta_{\mu\nu}+\frac{q_{\mu}q_{\nu}}{q^2}(\xi-1)\biggr),
\end{equation}%
where $\xi$ is the Fadeev-Popov gauge fixing parameter. $\xi=1$ corresponds to the Feynman gauge and $\xi=0$ to the Landau gauge. The $|q|$ dependence comes from calculating the self energy
\\
\\
\begin{equation} \label{EQ:f-bubble}
\parbox{15mm}{%30mm
\begin{fmfgraph*}(10,6)%(10,6)
\fmfleft{i}
\fmfright{o}
\fmf{plain_arrow,label=$k+q$,right}{o,i}
\fmf{plain_arrow,label=$k$,right}{i,o}
\fmfdot{i}
\fmfdot{o}
\end{fmfgraph*}
}.
\end{equation}
\\
\\
The charge $e$ has disappeared form the final result. This charge appears if one includes the $q^2$ dependence in the denominator of the gauge propagator. But in the small $q$ (long distance) limit, the $|q|$ dependence dominates over $q^2$ term, no matter how large $e$ is. Another way to say this is that the QED$_{3}$ is self--critical. It will flow to a stable fixed point ---even if we start with a large gauge charge $e$--- at least for some large $N$.

The gauge propagator of our theory, \textit{in the leading order}, will come \textit{solely} from the f-fermion bubble, given diagrammatically in Eq.~(\ref{EQ:f-bubble}). The $b$ propagation does not contribute to  the gauge field's self energy in the leading order.\ Therefore, in the leading order, $\left<a_{\mu}a_{\nu}\right>(q)$ of our Lagrangian density Eq.~(\ref{EQ:L_E}) is the same as QED$_{3}$.\\

\textit{The Kondo Field $b$}. The $b$ field propagator, in the leading order, is given by:
\\
\begin{equation}
\left<b b^*\right>^{-1}(q) \equiv G_b^{-1}(q)=
G_{b0}^{-1}(q)-
\parbox{15mm}{%30mm
\begin{fmfgraph*}(10,6)%(10,6)
\fmfleft{i}
\fmfright{o}
\fmf{dbl_plain_arrow,label=$k+q$,right}{o,i}
\fmf{plain_arrow,label=$k$,right}{i,o}
\end{fmfgraph*}
} ,\\
\\
\\
\end{equation}
\\
\\
where $G_{b0}^{-1}(q)=N/J_K + \mathcal{O}(q^2)$. The self energy is easily obtained to be:
\\
\begin{equation} \label{EQ:b_self_energy}
\begin{split}
\parbox{15mm}{%30mm
\begin{fmfgraph*}(10,6)%(10,6)
\fmfleft{i}
\fmfright{o}
\fmf{dbl_plain_arrow,label=$k+q$,right}{o,i}
\fmf{plain_arrow,label=$k$,right}{i,o}
\end{fmfgraph*}
} &= - (-i)^2 N \int\frac{d^d k}{(2\pi)^d} {\rm{Tr}}
	\biggl(\slashed{k}^{-1}\left(\slashed{k}+\slashed{q}\right)^{-1}\biggr) \\
&= - \frac{N}{4} |q| + \frac{2N \Lambda }{\pi^2} 
\end{split}
\end{equation}
This will result in the following $b$ propagator:
\begin{equation} \label{EQ:massive_b}
\left<b b^*\right>(q) =
\parbox{25mm}
{
\begin{fmfgraph*}(24,12)
\fmfstraight
\fmfleft{i}
\fmfright{o}
\fmf{dbl_dashes_arrow,label=$q$}{o,i}
\end{fmfgraph*}
} \ =\frac{4}{N\left(|q|+m_b\right)}\ , \\
\end{equation}
where $m_b$ is the \textit{effective mass} of the $b$ propagator, and is given by:
\begin{equation}
m_b=\frac{4}{J_K}-\frac{8\Lambda}{\pi^2}\propto (J_K^{cr}-J_K),
\end{equation} 
where $J_K^{cr}$  is given by $J_K^{cr}=\pi^2/(2\Lambda)$. $J_K^{cr}$ is of course nonuniversal, i.e. it depends on how we regularize the integral in Eq.~(\ref{EQ:b_self_energy}). However the final result that 
\begin{equation} \label{EQ:m:J-Jc}
m_b \propto (J_K^{cr}-J_K)
\end{equation}
is scheme independent. 

The above result, that the mass of the $b$ field can be tuned to \textit{zero}, serves as a sanity check for the mean-field result of the finite coupling Kondo transition obtained in Sec.~\ref{SEC:MFT}. 

Since the self energy has $|q|$ dependence, in the long distance (small $|q|$) limit, it dominates over the $q^2$ dependence coming from $|\partial_{\mu} b|^2$. The $q^2$ dependence is ignored in the final expression given in Eq.~(\ref{EQ:massive_b}). In the RG language, the $|\partial_{\mu} b|^2$ term is irrelevant at the fixed point.

We also mention a further notation for the \textit{massless} $b$ propagator. We have used the  ``double-dashed" line notation for $b$ propagator away from the critical point. We use a ``simple-dashed" line for the $b$ propagator at the critical point:
\begin{equation} \label{EQ:massless_b}
\parbox{20mm}
{
\begin{fmfgraph*}(20,12)
\fmfstraight
\fmfleft{i}
\fmfright{o}
\fmf{dashes_arrow,label=$q$}{o,i}
\end{fmfgraph*}
}=\frac{4}{N|q|}\ .
\end{equation}
Except for the calculation of $\gamma_b$, done in Sec.~\ref{APP:gamma_b}, massless $b$ propagator of Eq.~(\ref{EQ:massless_b}) is used throughout the paper.

The $b$ propagator given by Eqs.~(\ref{EQ:massive_b}) and~(\ref{EQ:massless_b}) is gauge invariant. This will not be the case to all orders in perturbation theory. The reason is as follows. There is a gauge freedom in defining $\Phi$. Given that the action is gauge invariant, the interaction $b\bar{\Psi}^a \Phi^a + b^*\bar{\Phi}^a \Psi^a$ dictates that gauge transformation
\begin{equation}
\Phi(x) \rightarrow e^{i\alpha(x)} \Phi(x)
\end{equation}
has to be accompanied by 
\begin{equation}
b(x) \rightarrow e^{-i\alpha(x)} b(x),
\end{equation}
since $\Psi$ can not tolerate any gauge freedom. So the $b$ propagator in principle should have a gauge dependence. However to \textit{leading order}, the gauge field propagator does not enter in $b$'s self energy and that is why the leading order $b$ propagator given by Eq.~(\ref{EQ:massive_b}) is gauge invariant. 

%***********************************************
\section{Critical Properties of the Fixed Point}
%***********************************************

%**************************
\subsection{Divergence of The Correlation Length: the Exponent $\nu$ \label{SEC:nu}}
%**************************
In this section we go through the calculation of the divergence of the correlation length $\xi$ near the fixed point by calculating the exponent $\nu$:
\begin{equation} 
	\xi \propto \frac{1}{|J_K-J_K^c|^{\nu}}.
	\label{nuDEF}
\end{equation}  

We mention first what the correlation length means by looking at the Kondo--Heisenberg Hamiltonian in different regimes. At $J_K = 0$ the Kondo--Heisenberg Hamiltonian has two decoupled parts: the kinetic term for the conduction electrons and the anti-ferromagnetic Heisenberg exchange of the localized spins. There is no entanglement between conduction electrons and the localized spins in the ground state. In the large $J_K$ limit the Kondo term dominates over the Heisenberg exchange. The ground state is the product of conduction electrons and the localized spin singlets at each site. Since we are at half filling every conduction electron is taken away by a localized spin. There is entanglement in the ground state. But the entanglement is restricted to one site. As we approach the critical point, the ``size'' of these Kondo singlets grows and it eventually diverges at the critical point. The scale associated with these Kondo singlets is what we mean by the correlation length $\xi$. 

The relevant flow of the mass term $|b|^2$ will introduce a diverging length scale $\xi_b$ near the transition point. Since the Kondo phase is charactarized by the condensation of the $b$ field, the correlation length $\xi$ is proportional to  $\xi_b$. This results in
\begin{equation} \label{EQ:nu=nu_b}
\nu=\nu_b.
\end{equation} 

We obtain $\nu_b$ by using the scaling relation\cite{Ma76}
\begin{equation}
\label{EQ:nuDef}
\nu_b=\frac{\gamma_b}{2-\eta_b},
\end{equation}
where $\gamma_b$ and $2-\eta_b$ characterize the power-law behavior of two-point function $G_b(k,m_b)$
\begin{equation}
\left<b(x) b^*(x')\right> = \int \frac{d^3 k}{(2\pi)^3} e^{ik\cdot(x-x')}G_b(k,m_b)
\end{equation}
in two different limits:
\begin{align}
\label{EQ:def_gammab} G_b(0,m_b\rightarrow 0) &\propto m_b^{-\gamma_b},\\
\label{EQ:def_etab}   G_b(k\rightarrow 0,0)  &\propto |k|^{-2+\eta_b}. 
\end{align}

It should be noted that since $b$ is not gauge invariant, $\gamma_b$ and $\eta_b$ are not gauge invariant. However $\nu$ by definition, given by Eq.~(\ref{nuDEF}), is a gauge invariant quantity. Since it is not a priori obvious that the scaling relation given by Eq.~(\ref{EQ:nuDef}) is gauge invariant, we calculate $\gamma_b$ and $\eta_b$ in a general gauge, and show that $\nu$ derived from Eq.~(\ref{EQ:nuDef}) is indeed gauge invaraint.

The methods we use to calculate $\gamma_b$ and $\eta_b$ are very different. The calculation of $\gamma_b$ is done by directly calculating the Green's function $G_b(0,m_b)$. Calculating $\eta_b$, in this manner, is much more involved.  Instead, we calculate $\eta_b$  by relating $\eta_b$ to the scaling dimension of $\bar{\Psi}^a\Phi^a$. The scaling dimension of $\bar{\Psi}^a\Phi^a$ is then obtained by perturbing the Lagrangian density $\mathcal{L}_E \rightarrow \mathcal{L}_E + \delta \bar{\Psi}^a \Phi^a$ and applying the  Callan-Symanzik equation  for the propagator $\left<\Psi^a\bar{\Phi}^a\right>$, in the presence of the perturbation, to obtain the flow of the composite operator $\bar{\Psi}^a \Phi^a$, and thus extracting its scaling dimension.

%*****************************************
\subsubsection{$\gamma_b$\label{APP:gamma_b}}
%*****************************************
To calculate $\gamma_b$ we calculate $G_b(0,m_b)$, i.e. the uniform part of the $b$ propagator  slightly away from the critical point. From the calculations in the previous section leading to  Eq.~(\ref{EQ:massive_b}) we have
\begin{equation}
G_b^{-1}(0,m_b) = N\frac{m_b}{4} - \Sigma_b^{1/N}(0,m_b),
\end{equation}
where $ \Sigma_b^{1/N}(0,m_b)$ is given diagramatically by
\begin{equation} \label{EQ:sigma_b(0,m)}
\begin{split}
\\
 \Sigma_b^{1/N}(0,m_b) &=
\parbox{41mm}
{
\begin{fmfgraph*}(40,15)
  \fmfleft{l}
  \fmfright{r}
  \fmf{dbl_plain_arrow,right=0.6,tension=1}{r,l}
  \fmf{plain_arrow}{l,p1,p2,r}
  \fmf{photon,left,tension=0}{p1,p2}
\end{fmfgraph*}
} \\
&+
\parbox{41mm}
{
\begin{fmfgraph*}(40,15)
  \fmfleft{l}
  \fmfright{r}
  \fmf{dbl_plain_arrow,right=0.6,tension=1}{r,l}
  \fmf{plain_arrow}{l,p1}
  \fmf{dbl_plain_arrow}{p1,p2}
  \fmf{plain_arrow}{p2,r}
  \fmf{dbl_dashes_arrow,right,tension=0}{p2,p1}
\end{fmfgraph*}
} \\
&+
\parbox{41mm}
{
\begin{fmfgraph*}(40,15)
  \fmfleft{l}
  \fmfright{r}
  \fmf{plain_arrow,left=0.6,tension=1}{l,r}
  \fmf{dbl_plain_arrow}{r,p2}
  \fmf{plain_arrow}{p2,p1}
  \fmf{dbl_plain_arrow}{p1,l}
  \fmf{dbl_dashes_arrow,right,tension=0}{p2,p1}
\end{fmfgraph*}
} \\
&+\parbox{41mm}
{
\begin{fmfgraph*}(40,15)
  \fmfleft{l}
  \fmfright{r}
  \fmfpolyn{phantom,tension=-0.5}{t}{3}
  \fmf{phantom,tension=25}{l,t1}
  \fmf{plain_arrow}{t1,t2}
  \fmf{plain_arrow}{t2,t3}
  \fmf{dbl_plain_arrow}{t3,t1}
  \fmfpolyn{phantom,tension=-0.5}{u}{3}
  \fmf{phantom,tension=25}{r,u1}
  \fmf{dbl_plain_arrow}{u1,u2}
  \fmf{plain_arrow}{u2,u3}
  \fmf{plain_arrow}{u3,u1}
  \fmf{photon}{u3,t2}
  \fmf{dbl_dashes_arrow}{u2,t3}
  \fmfdot{t2,u3}
\end{fmfgraph*}
} \\
\\
&= \frac{\xi-1}{\pi^2} m_b\log (m_b/\mu). \\
\end{split}
\end{equation}
This results in 
\begin{equation}
\begin{split}
G_b^{-1}(0,m_b)&= N \frac{m_b}{4} \biggl(1-\frac{4\left(\xi-1\right)}{N\pi^2} \log(m_b/\mu) \biggr) \\
&\propto m_b^{1-\frac{4\left(\xi-1\right)}{\pi^2 N}}, \\
\end{split}
\end{equation}
from which we obtain
\begin{equation}
\gamma_b = 1-\frac{4\left(\xi-1\right)}{\pi^2 N} .
\label{EQ:gamma_b}
\end{equation}

The integrals are evaluated in the dimensional regularization scheme.\cite{Brown,Collins} The mass scale $\mu$ in this scheme is introduced to keep track of dimensions in $d$ dimensional integration and plays the role of a UV cutoff. A \textit{simple pole} in the dimension $d$ reflects the \textit{logarithmic divergence} of the original 3 dimensional integration. We only keep track of the IR divergent parts. %@@@
As far as the universal properties of the fixed point are concerned this regularization would give the same answer as a cutoff regularization.

We briefly discuss the diagrams involved in $\Sigma_b^{1/N}(0,m_b)$. The first diagram of Eq.~(\ref{EQ:sigma_b(0,m)}) has no $m_b$ dependence. The second and third diagrams are equal and are given by:
\vspace{10mm}
\begin{equation}
\parbox{41mm}
{
\begin{fmfgraph*}(40,15)
  \fmfleft{l}
  \fmfright{r}
  \fmf{dbl_plain_arrow,right=0.7,tension=1,label=$p$}{r,l}
  \fmf{plain_arrow,label=$p$}{l,p1}
  \fmf{dbl_plain_arrow,label=$p+q$}{p1,p2}
  \fmf{plain_arrow,label=$p$}{p2,r}
  \fmf{dbl_dashes_arrow,right,tension=0,label=$q$}{p2,p1}
\end{fmfgraph*}
}=-\frac{1}{2\pi^2} m_b \log(m_b/\mu).
\end{equation}
The last diagram of Eq.~(\ref{EQ:sigma_b(0,m)}) contains the vertex that couples $b$ bosons with the gauge field:
\begin{equation}
\begin{split}
&\parbox{41mm}{
\begin{fmfgraph*}(40,40)
 \fmfsurround{l,r,t}
 \fmfpolyn{phantom,tension=-0.5}{c}{3}
 \fmf{dashes_arrow}{l,c1}
 \fmf{photon}{r,c2}
 \fmf{dashes_arrow}{c3,t}
 \fmf{dbl_plain_arrow,label=$p+k$,label.side=left}{c1,c3}
 \fmf{plain_arrow,label=$p-q$,label.side=left}{c3,c2}
 \fmf{plain_arrow,label=$p$,label.side=left}{c2,c1}
 \fmfv{decor.shape=circle,decor.filled=full,
      decor.size=2thick,label=$\mu$,label.dist=2mm,label.angle=180}{c2}
\end{fmfgraph*}
} \\
&=%(-1)(-i)^3 i 
+ N \int\frac{d^3 p}{(2\pi)^3} 
{\rm{Tr}}\biggl(\left(\slashed{p}+\slashed{k}\right)^{-1}\slashed{p}^{-1}\gamma^{\mu}\left(\slashed{p}-\slashed{q}\right)^{-1}\biggr).\\
\end{split}
\end{equation}
In the calculation of $\Sigma_b^{1/N}(0,m)$ the momentum entering this vertex is zero. The result of this integral for $k=0$ is 
\begin{equation}
\int\frac{d^3 p}{(2\pi)^3} {\rm{Tr}}\biggl(\slashed{p}^{-1} \slashed{p}^{-1}\gamma^{\mu}\left(\slashed{p}-\slashed{q}\right)^{-1}\biggr) = \frac{-q_{\mu}}{4|q|}.
\end{equation}
Having the result of the above integral, we get
\\
\begin{equation}
\sum_{\nu \sigma} \hspace{5mm} \parbox{41mm}
{
\begin{fmfgraph*}(40,15)
  \fmfleft{l}
  \fmfright{r}
  \fmfpolyn{phantom,tension=-0.5}{t}{3}
  \fmf{phantom,tension=25}{l,t1}
  \fmf{plain_arrow,label=$p_1+q$}{t1,t2}
  \fmf{plain_arrow,label=$p_1$}{t2,t3}
  \fmf{dbl_plain_arrow,label=$p_1+q$}{t3,t1}
  \fmfpolyn{phantom,tension=-0.5}{u}{3}
  \fmf{phantom,tension=25}{r,u1}
  \fmf{dbl_plain_arrow,label=$p_2+q$}{u1,u2}
  \fmf{plain_arrow,label=$p_2$,label.side=left}{u2,u3}
  \fmf{plain_arrow,label=$p_2+q$,label.side=right}{u3,u1}
  \fmf{photon,label=$q$,label.side=left}{u3,t2}
  \fmf{dbl_dashes_arrow,label=$q$}{u2,t3}
 \fmfv{decor.shape=circle,decor.filled=full,
      decor.size=2thick,label=$\nu$,label.dist=2mm}{t2}
 \fmfv{decor.shape=circle,decor.filled=full,
      decor.size=2thick,label=$\sigma$,label.dist=2mm}{u3}
%  \fmfdot{t2,u3}
\end{fmfgraph*}
}=\frac{\xi}{\pi^2} m_b \log(m_b/\mu).
\end{equation}
%*************************************
\subsubsection{$\eta_b$\label{APP:eta_b}}
%*************************************
$\eta_b$ is easily related to the scaling dimension of the $b$ field at the critical point [See Eq.~(\ref{EQ:def_etab})] :
\begin{equation}   \label{EQ:etab=-1+2b}
\eta_b=-1+2[b].
\end{equation}
Since the only way for the $b$ field to propagate is to decay into $\bar{\Psi}^a \Phi^a$, the scaling of the \textit{self energy} of $b$ is the same as the \textit{propagator} of the composite operator $\bar{\Psi}^a \Phi^a$. This observation results in
\begin{equation} \label{EQ:[b]=3-[ps]}
[b]=3-[\bar{\Psi}^a \Phi^a] .
\end{equation}
Thus $\eta_b$ is known once we know the scaling dimension of the composite operator $\bar{\Psi}^a \Phi^a$.

The scaling dimension of $\bar{\Psi}^a \Phi^a$ is obtained  by adding the perturbation
\begin{equation} \label{EQ:eta_vertex}
\mathcal{L}_E \rightarrow \mathcal{L}_E + \delta \bar{\Psi}^a \Phi^a
\end{equation}
to the Lagrangian density and obtaining the flow of $\delta$ by studying the Callan-Symanzik equation for the propagator $\left<\Psi^a(x) \bar{\Phi}^a(x')\right>$.

The Callan-Symanzik equation is essentially the ``scale invariance'' relation.\cite{Brown, Coleman} Near the fixed point, a change in scale has to be accompanied by the flow of some coupling constants and scaling of the fields to maintain scale invariance. For the propagator under consideration it will read
\begin{equation} \label{EQ:cs_psi_phi}
\biggl(-\frac{\partial}{\partial\ell} - \Delta_{\Psi}-\Delta_{\Phi}+ \frac{d C_{m}}{d\ell}\frac{\partial}{\partial C_m} \biggr)
  \left<\Psi^a(e^{\ell}x) \bar{\Phi}^a(e^{\ell}x')\right>=0,
\end{equation}
where $\Delta_{\Phi}$ and $\Delta_{\Psi}$ are scaling dimensions of the fields $\Phi$ and $\Psi$. The set C includes the coupling constants that flow. For the perturbation given by Eq.~(\ref{EQ:eta_vertex}) it has only 1 element:
\footnote{$\bar{\Psi}^a \Phi^a$ and  $\bar{\Phi}^a \Psi^a$ have different gauge charges and they do not mix together.}
\begin{equation}
C=\{\delta\}.
\end{equation}

It is more convenient to apply the Callan-Symanzik equation to the momentum-space Green's function $\left<\bar{\Psi}^a \Phi^a\right>(k)$
\begin{equation}
\left<\Psi^a \bar{\Phi}^a\right>(k)=\int \frac{d^3 k}{(2\pi)^3} e^{ik(x-x')}\left<\Psi^a(x) \bar{\Phi}^a(x')\right>.
\label{EQ:psiphi(k)}
\end{equation}
Eq.~(\ref{EQ:cs_psi_phi}) results in 
\begin{equation} \label{EQ:cs_psi_phi(k)}
\biggl(-\frac{\partial}{\partial\ell} +\left(3- \Delta_{\Psi}-\Delta_{\Phi}\right)+ \frac{d \delta}{d\ell}\frac{\partial}{\partial \delta}\biggr)\left<\bar{\Psi}^a \Phi^a\right>(e^{-\ell}k)=0,
\end{equation}
where ``3'' is coming from the scaling of the measure $d^3 k$ in the integrand of Eq.~(\ref{EQ:psiphi(k)}).

To apply the Callan-Symanzik equation we first need to obtain $\Delta_{\Phi}$ and $\Delta_{\Psi}$. The gauge dependant $\Delta_{\Phi}$ is obtained by calculating the propagator
\begin{equation}
\begin{split}
\left< \Phi^a \bar{\Phi}^a\right> &=
\parbox{41mm}%29
{
  \begin{fmfgraph}(40,18)
    \fmfstraight
    \fmfleft{i}
    \fmfright{o}
    \fmf{plain_arrow,label=$k$}{o,i}
 \end{fmfgraph}
} \\
&+\parbox{41mm}%36
{
\begin{fmfgraph}(40,18)
  \fmfleft{l}
  \fmfright{r}
  \fmf{plain_arrow}{r,c1}
  \fmf{plain_arrow,tension=-0.2}{c1,c2}
  \fmf{plain_arrow}{c2,l}  
  \fmf{photon,right}{c1,c2}
  \fmfdot{c1,c2}
\end{fmfgraph}
} \\
&+ \parbox{41mm}%36
{
\begin{fmfgraph}(40,18)
  \fmfleft{l}
  \fmfright{r}
  \fmf{plain_arrow}{r,c1}
  \fmf{dbl_plain_arrow,tension=-0.2}{c1,c2}
  \fmf{plain_arrow}{c2,l}  
  \fmf{dashes_arrow,left}{c2,c1}
\end{fmfgraph}
}.\\
\end{split}
\end{equation}
The result is
\begin{equation} 
\left< \Phi^a \bar{\Phi}^a\right>(k) = \left(i\slashed{k}\right)^{-1} \Biggl(1+ \frac{1}{N}\biggl(\frac{2}{\pi^2}+\frac{4}{\pi^2}\left(\xi-1\right)\biggr) \log(|k|/\mu) \Biggr), 
\end{equation}
from which we get
\begin{equation}
\Delta_{\Phi}=1+\frac{1}{N}\Bigl(\frac{1}{\pi^2}+\frac{2}{\pi^2}\left(\xi-1\right)\Bigr).
\end{equation}
Similarly, calculating the \textit{gauge invariant} conduction electron propagator 
\begin{equation} \label{EQ:psi-propagator}
\begin{split}
\left< \Psi^a \bar{\Psi}^a\right> &=
\parbox{41mm}%29
{
   \begin{fmfgraph}(40,18)
     \fmfstraight
     \fmfleft{i}
     \fmfright{o}
     \fmf{dbl_plain_arrow}{o,i}
   \end{fmfgraph}
} \\
&+ \parbox{41mm}%36
{
   \begin{fmfgraph}(40,18)
     \fmfleft{l}
     \fmfright{r}
     \fmf{dbl_plain_arrow}{r,c1}
     \fmf{plain_arrow,tension=-0.2}{c1,c2}
     \fmf{dbl_plain_arrow}{c2,l}  
     \fmf{dashes_arrow,right}{c1,c2}
   \end{fmfgraph}
 } \\
&= (i\slashed{k})^{-1} \biggl(1+\frac{2}{3\pi^2N}\log(|k|/\mu)\biggr) \\
\end{split}
\end{equation}
results in the gauge invariant $\Delta_{\Psi}$
\begin{equation}
\Delta_{\Psi} = 1+\frac{1}{3 \pi^2 N}.
\end{equation}

The logarithmic infrared divergence of Eq.~(\ref{EQ:psi-propagator}) can be interpreted as the scaling of the quasi-particle weight $Z$, defined by $ \left< \Psi^a \bar{\Psi}^a\right>(k)=Z(i\slashed{k})^{-1} $ . The scaling form  of $Z$ which is consistent with Eq.~(\ref{EQ:psi-propagator}) is 
\begin{equation}
Z=|\frac{k}{\mu}|^{\frac{2}{3\pi^2 N}}.
\end{equation}
The vanishing of the $Z$ factor as $k\rightarrow 0$ can in principle be seen in tunnelling experiments. This non-Fermi liquid behavior arises, because at the KI--ASL fixed point, there are very strong scattering of conduction electrons into the f-fermions. This is due to the fact that the $b$ field which causes the scattering has become massless.

The next step is to obtain $\left< \bar{\Psi}^a \Phi^a\right>(k)$ in the presense of the perturbation $\delta \bar{\Psi}^a \Phi^a$ given by Eq.~(\ref{EQ:eta_vertex}). This perturbation is represented diagrammatically by
\begin{equation}
  \parbox{33mm}{
  \begin{fmfgraph*}(30,18)
    \fmfstraight
    \fmfleft{l}
    \fmfright{r}
    \fmf{plain_arrow}{r,o}
    \fmf{dbl_plain_arrow}{o,l}
    \fmfv{decor.shape=hexagram,decor.filled=0,decor.size=4mm}{o}
  \end{fmfgraph*}
} = \delta.
\end{equation}
The diagrammatic expression for $\left<\Psi^a \bar{\Phi}^a\right>$ is then given by
\begin{equation}
\begin{split}
\left<\Psi^a \bar{\Phi}^a\right> &=
\parbox{45mm}{
  \begin{fmfgraph*}(44,18)
    \fmfstraight
    \fmfleft{l}
    \fmfright{r}
    \fmf{plain_arrow}{r,o}
    \fmf{dbl_plain_arrow}{o,l}
    \fmfv{decor.shape=hexagram,decor.filled=0,decor.size=4mm}{o}
  \end{fmfgraph*}
} \\
&+\parbox{45mm}
{
\begin{fmfgraph*}(44,18)
  \fmfleft{l}
  \fmfright{r}
  \fmf{plain_arrow}{r,c1}
  \fmf{plain_arrow,tension=-0.2}{c1,c2}
  \fmf{plain_arrow}{c2,c3}
  \fmf{dbl_plain_arrow}{c3,l}
  \fmfv{decor.shape=hexagram,decor.filled=0,decor.size=4mm}{c3}  
  \fmf{photon,left}{c2,c1}
  \fmfdot{c1,c2}
\end{fmfgraph*}
} \\
&+\parbox{45mm}
{
\begin{fmfgraph*}(44,18)
  \fmfleft{l}
  \fmfright{r}
  \fmf{plain_arrow}{r,c1}
  \fmf{dbl_plain_arrow,tension=-0.2}{c1,c2}
  \fmf{plain_arrow}{c2,c3}
  \fmf{dbl_plain_arrow}{c3,l}
  \fmfv{decor.shape=hexagram,decor.filled=empty,decor.size=4mm}{c3}  
  \fmf{dashes_arrow,left}{c2,c1}
\end{fmfgraph*}
} \\
&+\parbox{45mm}
{
  \begin{fmfgraph*}(44,18)
    \fmfleft{l}
    \fmfright{r}
    \fmf{plain_arrow}{r,c1}
    \fmf{dbl_plain_arrow}{c1,c2}
    \fmf{plain_arrow,tension=-0.2}{c2,c3}
    \fmf{dbl_plain_arrow}{c3,l}
    \fmfv{decor.shape=hexagram,decor.filled=empty,decor.size=4mm}{c1}  
    \fmf{dashes_arrow,right}{c2,c3}
  \end{fmfgraph*}
} \\
&+
\parbox{45mm}
{
\begin{fmfgraph*}(44,44)
  \fmfsurroundn{f}{3}
  \fmfleft{f1}
  \fmftop{f2}
  \fmfright{f3}
  \fmfpolyn{phantom,tension=-0.5}{i}{3}
  \fmf{plain_arrow,tension=4}{f3,r}
  \fmf{plain_arrow,tension=6}{r,l}
  \fmf{dbl_plain_arrow,tension=4}{l,f1}
  \fmf{phantom,tension=20}{i1,f2}
  \fmf{dbl_plain_arrow}{i1,i2}
  \fmf{plain_arrow}{i2,i3}
  \fmf{plain_arrow}{i3,i1}
  \fmf{photon,tension=1}{r,i3}
  \fmf{dashes_arrow,tension=1}{i2,l}
  \fmfv{decor.shape=hexagram,decor.filled=empty,decor.size=4mm}{i1}  
  \fmfdot{i3,r}
\end{fmfgraph*}
}. \\
\end{split}
\end{equation}
It results in 
\begin{equation} \left<\Psi^a \bar{\Phi}^a\right>(k) =\delta \ \left(i\slashed{k}\right)^{-2} \biggl( 1- \frac{4}{3\pi^2N}\log(|k|/\mu)\biggr) .\end{equation}
Callan-Symanzik equation [Eq.~(\ref{EQ:cs_psi_phi(k)})] applied to the above propgator resutls in
\begin{equation}
\frac{d\delta}{d l} = \biggl(1+\frac{8}{3\pi^2 N }+\frac{2}{\pi^2 N}\left(\xi-1\right)\biggr)\delta.
\end{equation}
The above relation together with Eq.~(\ref{EQ:[b]=3-[ps]}) yields:
\begin{equation}
[b]=1+\frac{1}{N}\biggl(\frac{8}{3\pi^2}+\frac{2}{\pi^2}\left(\xi-1\right) \biggr).
\end{equation}
Combine this with Eq.~(\ref{EQ:etab=-1+2b}) to get
\begin{equation}
\eta_b = 1 + \frac{1}{N}\biggl(\frac{16}{3\pi^2}+\frac{4}{\pi^2}\left(\xi-1\right)\biggr).
\label{EQ:eta_b}
\end{equation}

Collecting what we have obtained for $\gamma_b$ and $\eta_b$ given by Eqs.~(\ref{EQ:gamma_b}) and~(\ref{EQ:eta_b}), together with the scaling relation given by Eq.~(\ref{EQ:nuDef}), we arrive at the gauge invariant 
\begin{equation}
\nu=1+\frac{16}{3\pi^2N}.
\end{equation}
Note that the $1/N$ fluctuations have brought us \textit{further away} from the generic $1/2+\mathcal{O}(\epsilon)$ result one would obtain for a stable fixed point in an $\epsilon$ expansion (which is an expansion around the gaussian action), assuming there exists such a stable fixed point.

%**************************************
\subsection{Staggered Spin Correlation \label{SEC:SSC}}
%**************************************

We continue the study of the physical properties of our KI--ASL fixed point by obtainning staggered spin $\bm{M}=\bm{S}_A-\bm{S}_B$ correlation \textit{at} the fixed point. The correlation is algebraic and our fixed point is an algebraic spin liquid fixed point. \

Since the decay exponent of  $\left<\bm{M}(x) \cdot \bm{M}(x')\right>$ and $\left<M^{+}(x) M^{-}(x')\right>$ is the same, it suffices to find the scaling dimension of $M^{+}$. In terms of the field operators $M^{+}(x)$ is
\begin{eqnarray}
M^+(x) &=& \bar{\Phi}^1(x) \mathcal{M} \Phi^2(x),
\end{eqnarray}
where $\mathcal{M}$ is $4\times4$ matrix, which should be obtained from the relation between the field $\Phi$ and the microscopic operators developed in Sec.~\ref{SEC:EFT}. Physically 1 and 2 represents $\uparrow$  and $\downarrow$. In $1/N$ expansion they are 2 indices among many. $\mathcal{M}$ is obtained to be \footnote{See the Appendix~\ref{APP:Staggered Spin}.}
\begin{equation}
 \mathcal{M} = \mathbf{1} \otimes \sigma_3 .
\end{equation}

We should add this perturbation to the Lagrangian and study its flow. However  $\bar{\Phi}^1(x) \mathcal{M} \Phi^2(x)$ mixes with $\bar{\Psi}^1(x) \mathcal{M} \Psi^2(x)$. In other words $M^+$ is not a scaling operator.
% However $\bar{\Phi}^1(x) \mathcal{M} \Phi^2(x)$ mixes with $\bar{\Psi}^1(x) \mathcal{M} \Psi^2(x)$. In another words $M^+$ is not a scaling operator. 
Having this in mind, we perturb the Lagrangian density by
\begin{equation} 
\delta \mathcal{L}_E= U_+\ \bar{\Phi}^1(x) \mathcal{M} \Phi^2(x) + u_+ \ \bar{\Psi}^1(x) \mathcal{M} \Psi^2(x)
\end{equation} %
and study the Callan-Symanzik equation for the propagators $\left<\Psi^1 \bar{\Psi}^2\right>(k)$ and $\left<\Phi^1 \bar{\Phi}^2\right>(k)$.
The result (obtained in Appendix~\ref{APP:Staggered Spin}) is
\begin{equation} \label{EQ:stagg_anam_matrix}
\frac{\partial}{\partial \ell}
\begin{bmatrix}
U_+\\
\\
u_+\\
\end{bmatrix}
=
\begin{bmatrix}
1+\frac{10}{\pi^2 N} & \frac{-2}{\pi^2 N}  \\
 & \\
\frac{-2}{\pi^2 N} & 1-\frac{2}{3 \pi^2 N} \\
\end{bmatrix}
\begin{bmatrix}
U_+\\
\\
u_+\\
\end{bmatrix}.
\end{equation}
We emphasize that off--diagonal terms make $1/N$ correction to the scaling dimension of the staggered spin. This is because the scaling dimensions of $\Psi$ and $\Phi$ are equal in the infinite $N$ limit. This is similar to the importance of off-diagonal terms in the 1st order \textit{degenerate} perturbation theory in quantum mechanics.

The largest eigenvalue $1+2\left(7+\sqrt{73}\right)/(3\pi^2 N)$ of above matrix dominates the scaling of $M^{+}$. The \textit{dominant} scaling dimension of $M^+$, denoted by $[M^{+}]$, is  then given by
\begin{equation}
[M^{+}]=2-\frac{2\left(7+\sqrt{73}\right)}{3\pi^2 N} \approx  2 -  \frac{31.1}{3\pi^2 N} .
\end{equation}
The decay exponent of the correlation function results immediately:
\begin{equation}
 \left<M(x)\cdot M(x')\right> \propto \frac{1}{|x-x'|^{2[M^+]}} + \cdots ,
\end{equation}
where the ellipsis denotes the decay with the larger exponent coming from the other eigenvalue of the matrix in Eq.~(\ref{EQ:stagg_anam_matrix}).

Finally, we mention a  comparison between our result and the Rantner--Wen studies.\cite{Rantner02} They studied spin correlations in their large-$N$ QED$_{3}$ effective field theory in the Landau gauge. They were interested in the similar bilinear form as we have here with $\mathcal{M}=\mathbf{1}$. In our calculation any bilinear form with the property $[\mathcal{M},\gamma_{\mu}]=0$ has the same scaling dimension. To check our result we turned off all contributions to the exponent coming from terms involving $\Psi$. The result $[M^{+}]_{{\rm ASL}}=2-32/(3\pi^2 N)$ obtained in Feynman gauge agrees with theirs obtained in Landau gauge. $2[M^{+}]_{{\rm ASL}}$ corresponds to the decay exponent of localized spin correlation when we are in the ASL phase, where the Kondo field is massive. As expected, we have $[M^{+}]_{{\rm ASL}} <  [M^{+}] $, i.e. the Neel fluctuations are stronger in the ASL phase than at the KI transition, where the Kondo singlets are forming.

In the $J<J_K^{cr}$ regime the ASL is in the presence of the semi-metal (described by $\Psi$ field) and the gapped Kondo field $b$. The question may arise whether the semi-metal would destabilize the ASL. The semi-metal will not destabilize the ASL, because integrating out $\Psi$ and $b$ fields in this phase will produce (retarded) four-fermion interaction term among $\Phi$ fields. These four-fermion interactions are irrelevant at the ASL fixed point.

%***********************************************************
\section{Stability of the Fixed Point \label{SEC:stability}}
%***********************************************************

In this section we discuss the issue of the stability of the KI--ASL fixed point in the honeycomb lattice to 1st order in $1/N$. There are two separate issues one has to consider in the stabillity of this fixed point.

In going to the continuum limit we have dropped the compact character of the gauge field. This is only justified if the instanton events, where the flux of the gague field suddenly changes by 2$\pi$, do not proliferate. For $J_K>J_K^{cr}$ the Kondo field $b$ is condensed and we are in the Higgs phase. Gauge field is massive, its fluctuations are supperessed and instantons do not proliferate. Addressing the proliferation of instantons in the regime where the Kondo phase has disappeared is more subtle. We know from Polyakov's work that the compact \textit{pure} gauge theory in 2+1 dimension is in the confining phase for all coupling constant, i.e. for all coupling constants the instantons proliferate.\cite{Polyakov} It has been argued recently that coupling the gauge field to massless Dirac fields results in a deconfined phase  for sufficiently large $N$.\cite{Hermele04} This is due to the fact that the massless matter field, in this case Dirac fields, suppress the fluctuations of the gauge field. We apply that argument here and assume that the gauge field in the $J_K< J_K^{cr}$ regime is in a deconfined phase.\ At the KI--ASL fixed point, due to the presence of  an extra massless matter field, i.e. the Kondo field $b$, the gauge field fluctuations are suppressed even more.
 
Next we focus on the instanton--free sector. At the fixed point the gauge field $a_{\mu}$, boson $b$ and two fermion fileds $\Psi$, $\Phi$; all have scaling dimension $1+\mathcal{O}(1/N)$. This can be seen by the form of propagators derived in Sec.~\ref{SEC:Feynman_Diagrams}. The large $N$ scaling dimension of $a_{\mu}$, $b$, $\Psi$ and $\Phi$ will limit our choices for relevant perturbations. Different such perturbations are discussed below:
\begin{itemize}
\item $b$ mass term: this term is already present in the microscopic bare Lagrangian. We have to \textit{tune} $J_K$ to make the effective mass of the $b$ field to be zero. This perturbation is indeed relevant.
\item $\Psi$ and $\Phi$ mass terms: these mass terms either break the SU(N) flavor symmetry or the particle-hole and rotation symmetry.
\item $b$ kinetic term of the form $b^*\left(\partial_{\mu}+ia_{\mu}\right)b$:  this term violates the particle--hole symmetry. It has to be accompanied by $b(\partial_{\mu}-ia_{\mu})b^*$ with an equal coefficient. But $b^*\partial_{\mu}b+b\partial_{\mu}b^*=\partial_{\mu}\left(b^* b \right)$ is a total derivative. 
\item $b$ kinetic term of the form $\delta \mathcal{L}=Ng|\left(\partial_{\mu}+ia_{\mu}\right)b|^2$: this term is irrelevant by the power counting. That is 
  \begin{equation}\nonumber 
  \frac{dg}{dl}=\bigl(-1+\mathcal{O}(1/N)\bigr)g.
  \end{equation}
\item Kinetic terms for $\Psi$ and $\Phi$ fields: these are the only potentially dangerous perturbations, because they are marginal in the $N\rightarrow \infty$ limit. Actually in the derivation of the Lagrangian density Eq.~(\ref{EQ:L_E}) we set $v_c=v_f$ to make the action Lorentz invariant. We have to consider perturbations that break this ``tuned'' Lorentz symmetry and study their flow. By scaling argument the flow of these perturbations is marginal to first order in $1/N$: marginally irrelevant, relevant, or exactly marginal. Our calculations results in the marginal irrelavance of these types of perturbations. We discuss this issue below.
\end{itemize}

We break the Lorentz symmetry by adding the perturbation
\begin{equation} \nonumber
\delta \mathcal{L}_E^a = \sum_{\sigma} \biggl( \delta_{\sigma}\  \bar{\Psi}^a \gamma_{\sigma} \partial_{\sigma} \Psi^a
+ \Delta_{\sigma} \ \bar{\Phi}^a \gamma_{\sigma} \left(\partial_{\sigma}-ia_{\sigma}\right) \Phi^a \biggr)
\end{equation}
to the Lagrangian density. The perturbation has to be considered both in $\Psi$ and $\Phi$ fields because of their mixing due to the exchange of the $b$ field. Applying Callan-Symanzik equation to the propagators $\left< \Psi^a \bar{\Psi}^a\right>$ and $\left< \Phi^a \bar{\Phi}^a\right>$ results in the following RG flow:
\begin{equation}
\frac{\partial}{\partial \ell}
\begin{bmatrix}
\Delta_0\\
\Delta_1\\
\Delta_2\\
\delta_0\\
\delta_1\\
\delta_2\\
\end{bmatrix}
= \mathcal{D}
\begin{bmatrix}
\Delta_0\\
\Delta_1\\
\Delta_2\\
\delta_0\\
\delta_1\\
\delta_2\\
\end{bmatrix} ,
\end{equation}
where the matrix anomalous dimension $\mathcal{D}$ is given by
\begin{equation}
\mathcal{D} =
\frac{1}{15\pi^2 N}
\begin{bmatrix}
-74 & 32 & 32 & -2 & -4 & -4 \\
32 & -74 & 32 & -4 & -2 & -4 \\
32 & 32 & -74 & -4 & -4 & -2\\
 -2 & -4 & -4 & -10 & 0 & 0 \\
 -4 & -2 & -4 & 0 & -10 & 0\\
 -4 & -4 & -2 & 0 & 0 & -10\\
\end{bmatrix} .
\end{equation}
Eigenvalues of this matrix anomalous dimension are all negative except for one trivial zero eigenvalue. The zero eigenvalue is trivial since the corresponding perturbation can be absorbed by scaling $\Psi$ and $\Phi$ fields, shown in Appendix~\ref{APP:Lorentz Restoration}. 

We knew from the mean-field result that there exist a fixed point in the parameter space. But our results goes further by ruling out a multi--critical fixed point. So based on our result, for large enough N, this quantum critical point will be achieved by tuning one parameter $J_K$.

\section{Conclusions}
We have studied  the Kondo--Heisenberg model on the honeycomb lattice at half filling. The main motivation for our study is that the Kondo band gap vanishes at a finite Kondo coupling constant (even as $J_H\rightarrow0$) so that a novel kind of quantum phase transition becomes possible.

In this paper we considered the transition of this Kondo insulator (KI) to a spin liquid phase which is later identified to be algebraic spin liquid (ASL). We developed a mean-field theory for such a transition using a fermionic representation for spins. To consider fluctuations about this mean-field systematically, we  adopted a large $N$ generalization of our model by having $N$ spin-flavors rather than just $\{\uparrow ,\downarrow\}$.  We found that the KI--ASL transition is controlled by a stable Lorentz invariant fixed point. 

In the KI phase, the Kondo field $b$ condenses and we are in the Higgs phase where the gauge field is massive. ASL phase is described by ``decoupled" QED$_{3}$ (coming from $J_H \bm{S}_i \cdot \bm{S}_j$ interaction) and the Dirac conduction electrons. They are ``decoupled" in that the Kondo field that couples these two terms is now massive. The gauge field is deconfined in the ASL.

We were able to calculate the exponent $\nu=1+16/(3\pi^2N)$ of the diverging length scale near the transition. Interestingly, $1/N$ fluctuations have pushed the infinite $N$ result further away from the result $\nu=1/2$ one would get from the Landau-type expansion. We noted that at our fixed point the quasi-particle weight of the conduction electron vanishes which is indicative of non-Fermi liquid behavior. We also calculated the decay exponent of the staggered spin $\bm{M}=\bm{S}_A-\bm{S}_B$ correlation at the fixed point. This is not a scaling operator, since it mixes with $\bm{m}=\bm{s}_A-\bm{s}_B$. The dominant decay exponent turned out to be $2[M^{+}]=4-4\left(7+\sqrt{73}\right)/(3\pi^2 N)\approx 4-62.2/(3\pi^2 N)$. On the other hand, in the ASL phase $\bm{M}$ is a scaling operator with the decay exponent $2[M^{+}]_{{\rm ASL} }=4-64/(3\pi^2N)$. The \textit{jump} in this exponent, as one gets to the KI transition point, is due to the fact that there exists an extra massless matter field at the transition (in this case Kondo field $b$), which modifies the exponent. This jump and its significance was discussed in a different context recently.\cite{YRan06}

The quantum phase transition between the KI and ASL does not break any physical symmetry. The KI--ASL transition is essentially a Higgs--deconfinement transition. In other words, what changes at the transition point is the dynamics of the gauge field. This by itself is an interesting property of the transition we have studied, which is realized in a minimal Kondo--Heisenberg model.

In connection to this work, we are currently pursuing the numerical study of this model; using a quantum Monte Carlo technique without the fermion sign problem.

\section{Acknowledgments}
We thank Michael Hermele for many insightful comments and suggestions and we thank T. Senthil and X.-G. Wen for helpful discussions. PAL acknowledges the support by the Department of Energy under grant DE-FG02-03ER46076.

\newpage
%///////////////////////////////////////////////////////////////////////////////////////////////////
\appendix
\section{Restoration of the Lorentz Invariance \label{APP:Lorentz Restoration}}
%******************************************************************************

Here we will give more details regarding the irrelavance of the perturbations that break the Lorentz symmetry at our Lorentz invariant KI--ASL fixed point. It is tempting to break the Lorentz invariance by perturbing only one field, $\Psi$ or $\Phi$. However  $\Psi$ and $\Phi$ loose their own identity as $b$ becomes massless and they mix together. Having this mixing in mind we consider the most general perturbation which is of the kinetic form for the $\Psi$ and $\Phi$ fields, and is consistent with the gauge symmetry 
\begin{equation} \label{EQ:break_lorentz_perturbation}
\delta \mathcal{L}_E^a = \sum_{\sigma} \biggl( \delta_{\sigma}\  \bar{\Psi}^a \gamma_{\sigma} \partial_{\sigma} \Psi^a
+ \Delta_{\sigma} \ \bar{\Phi}^a \gamma_{\sigma} \left(\partial_{\sigma}-ia_{\sigma}\right) \Phi^a \biggr) .
\end{equation}
%There is no summation over $a$ on the right hand side, and that is why the left hand side in the above notation has an index $a$. 
The corrections to fermion propagators and gauge vortex is represented diagrammatically:
\begin{equation}
\begin{split}
  \parbox{33mm}{
    \begin{fmfgraph*}(30,18)
      \fmfstraight
      \fmfleft{i}
      \fmfright{o}
      \fmf{dbl_plain_arrow,label=$q$}{o,c}
      \fmf{dbl_plain_arrow,label=$q$}{c,i}
      \fmfv{decor.shape=circle,decor.filled=empty,decor.size=4mm,label=$\sigma$,label.dist=-1mm}{c}
    \end{fmfgraph*}
  } &= -i\delta_{\sigma} \slashed{q}_{\sigma},\\
  \parbox{33mm}{
    \begin{fmfgraph*}(30,18)
      \fmfstraight
      \fmfleft{i}
      \fmfright{o}
      \fmf{plain_arrow,label=$q$}{o,c}
      \fmf{plain_arrow,label=$q$}{c,i}
      \fmfv{decor.shape=square,decor.filled=empty,decor.size=4mm,label=$\sigma$,label.dist=-1mm}{c}
    \end{fmfgraph*}
  } &= -i\Delta_{\sigma} \slashed{q}_{\sigma},\\
  \parbox{33mm}{
    \begin{fmfgraph*}(30,25)
      \fmfstraight
      \fmfright{i}
      %    \fmfright{i1,i2}
      \fmfleft{o}
      \fmftop{p}
      \fmf{plain_arrow}{i,c}
      %    \fmf{dbl_plain_arrow}{i1,c}
      \fmf{plain_arrow}{c,o}
      \fmf{photon}{c,p}
      %    \fmf{photon}{i2,c}
      \fmfv{decor.shape=square,decor.filled=empty,decor.size=4mm,label=$\sigma$,label.dist=-1mm}{c}
    \end{fmfgraph*}
  } &= +i\Delta_{\sigma} \gamma_{\sigma},
\end{split}
\end{equation}
where we have used the notation
\begin{equation} 
\slashed{q}_{\sigma} \equiv q_{\sigma}\gamma_{\sigma} .
\end{equation}

We evaluate the two propagators $\left< \Psi^a \bar{\Psi}^a\right>$ and $\left< \Phi^a \bar{\Phi}^a\right>$ in the presence of the perturbation given by Eq.~(\ref{EQ:break_lorentz_perturbation}). $\left< \Psi^a \bar{\Psi}^a\right>$ is given by
\begin{equation}
\begin{split}
  \left< \Psi^a \bar{\Psi}^a\right> &=
  \parbox{41mm}%29
	 {
	   \begin{fmfgraph}(40,18)
	     \fmfstraight
	     \fmfleft{i}
	     \fmfright{o}
	     \fmf{dbl_plain_arrow}{o,i}
	   \end{fmfgraph}
	 } \\
	 &+\parbox{41mm}%36
	 {
	   \begin{fmfgraph}(40,18)
	     \fmfleft{l}
	     \fmfright{r}
	     \fmf{dbl_plain_arrow}{r,c1}
	     \fmf{plain_arrow,tension=-0.2}{c1,c2}
	     \fmf{dbl_plain_arrow}{c2,l}  
	     \fmf{dashes_arrow,right}{c1,c2}
	   \end{fmfgraph}
	 } \\
	 &+\parbox{41mm}%28
	 {
	   \begin{fmfgraph*}(40,18)
	     \fmfstraight
	     \fmfleft{i}
	     \fmfright{o}
	     \fmf{dbl_plain_arrow}{o,c}
	     \fmf{dbl_plain_arrow}{c,i}
	     \fmfv{decor.shape=circle,decor.filled=empty,decor.size=4mm,label=$\sigma$,label.dist=-1mm}{c}
	   \end{fmfgraph*}
	 } \\
	 &+\parbox{41mm}
	 {
	   \begin{fmfgraph*}(40,18)
	     \fmfleft{l}
	     \fmfright{r}
	     \fmf{dbl_plain_arrow}{r,c1}
	     \fmf{plain_arrow,tension=-0.2}{c1,c2}
	     \fmf{dbl_plain_arrow}{c2,c3}
	     \fmf{dbl_plain_arrow}{c3,l}
	     \fmfv{decor.shape=circle,decor.filled=empty,decor.size=4mm,label=$\sigma$,label.dist=-1mm}{c3}  
	     \fmf{dashes_arrow,right}{c1,c2}
	   \end{fmfgraph*}
	 } \\
&+\parbox{41mm}
	 {
	   \begin{fmfgraph*}(40,18)
	     \fmfleft{l}
	     \fmfright{r}
	     \fmf{dbl_plain_arrow}{r,c1}
	     \fmf{dbl_plain_arrow}{c1,c2}
	     \fmf{plain_arrow,tension=-0.2}{c2,c3}
	     \fmf{dbl_plain_arrow}{c3,l}
	     \fmfv{decor.shape=circle,decor.filled=empty,decor.size=4mm,label=$\sigma$,label.dist=-1mm}{c1}  
	     \fmf{dashes_arrow,right}{c2,c3}
	   \end{fmfgraph*}
	 } \\
	 &+\parbox{41mm}
	 {
	   \begin{fmfgraph*}(40,18)
	     \fmfleft{l}
	     \fmfright{r}
	     \fmf{dbl_plain_arrow}{r,b1}
	     \fmf{plain_arrow,tension=-1}{b1,c}
	     \fmf{plain_arrow,tension=-1}{c,b2}
	     \fmf{dbl_plain_arrow}{b2,l}
	     \fmfv{decor.shape=square,decor.filled=empty,decor.size=4mm,label=$\sigma$,label.dist=-1mm}{c}  
	     \fmf{dashes_arrow,right}{b1,b2}
	   \end{fmfgraph*}
	 },
\end{split}
\end{equation}
where the sum over the space--time index $\sigma$, appearing in the above diagrams, is understood. Calculation of the diagrams --- done in Feynman gauge --- results in:
\<
\begin{split}
\left< \Psi^a \bar{\Psi}^a\right>(k) &= -i\slashed{k}^{-1} \biggl(1+\frac{2}{3\pi^2N}\log(|k|/\mu)\biggr) \\
 &+i\slashed{k}^{-1}\slashed{k}_{\sigma}\slashed{k}^{-1}\biggl(\delta_{\sigma}+\frac{\log(|k|/\mu)}{15\pi^2N} |\epsilon_{\sigma\mu\nu}|\  S_{\sigma;\mu\nu} \biggr), 
\end{split} 
\>
where $S_{\sigma;\mu\nu}$ is given by
\< S_{\sigma;\mu\nu} = 20 \delta_{\sigma} +2\Delta_{\sigma}+4\Delta_{\mu}+4\Delta_{\nu}.
\>
The above propagator has a contribution from the unperturbed fixed-point action, which gives the gauge independent  scaling dimension for $\Psi$ field
\<
\Delta_{\Psi} = 1+\frac{1}{3 \pi^2 N}.
\>
Next we calculate $\left< \Phi^a \bar{\Phi}^a\right>$ which is given by
\<
\begin{split}
  \left< \Phi^a \bar{\Phi}^a\right> &=
  \parbox{41mm}%29
	 {
	   \begin{fmfgraph}(40,18)
	     \fmfstraight
	     \fmfleft{i}
	     \fmfright{o}
	     \fmf{plain_arrow}{o,i}
	   \end{fmfgraph}
	 } \\
	 &+\parbox{41mm}%36
	 {
	   \begin{fmfgraph}(40,18)
	     \fmfleft{l}
	     \fmfright{r}
	     \fmf{plain_arrow}{r,c1}
	     \fmf{plain_arrow,tension=-0.2}{c1,c2}
	     \fmf{plain_arrow}{c2,l}  
	     \fmf{photon,right}{c1,c2}
	     \fmfdot{c1,c2}
	   \end{fmfgraph}
	 } \\
	 &+\parbox{41mm}%36
	 {
	   \begin{fmfgraph}(40,18)
	     \fmfleft{l}
	     \fmfright{r}
	     \fmf{plain_arrow}{r,c1}
	     \fmf{dbl_plain_arrow,tension=-0.2}{c1,c2}
	     \fmf{plain_arrow}{c2,l}  
	     \fmf{dashes_arrow,left}{c2,c1}
	   \end{fmfgraph}
	 } \\
	 &+\parbox{41mm}%28
	 {
	   \begin{fmfgraph*}(40,18)
	     \fmfstraight
	     \fmfleft{i}
	     \fmfright{o}
	     \fmf{plain_arrow}{o,c}
	     \fmf{plain_arrow}{c,i}
	     \fmfv{decor.shape=square,decor.filled=empty,decor.size=4mm,label=$\sigma$,label.dist=-1mm}{c}
	   \end{fmfgraph*}
	 } \\
	 &+\parbox{41mm}
	 {
	   \begin{fmfgraph*}(40,18)
	     \fmfleft{l}
	     \fmfright{r}
	     \fmf{plain_arrow}{r,c1}
	     \fmf{plain_arrow,tension=-0.2}{c1,c2}
	     \fmf{plain_arrow}{c2,c3}
	     \fmf{plain_arrow}{c3,l}
	     \fmfv{decor.shape=square,decor.filled=empty,decor.size=4mm,label=$\sigma$,label.dist=-1mm}{c3}  
	     \fmf{photon,left}{c2,c1}
	     \fmfdot{c1,c2}
	   \end{fmfgraph*}
	 } \\
	 &+\parbox{41mm}
	 {
	   \begin{fmfgraph*}(40,18)
	     \fmfleft{l}
	     \fmfright{r}
	     \fmf{plain_arrow}{r,c1}
	     \fmf{plain_arrow}{c1,c2}
	     \fmf{plain_arrow,tension=-0.2}{c2,c3}
	     \fmf{plain_arrow}{c3,l}
	     \fmfv{decor.shape=square,decor.filled=empty,decor.size=4mm,label=$\sigma$,label.dist=-1mm}{c1}  
	     \fmf{photon,left}{c3,c2}
	     \fmfdot{c3,c2}
	   \end{fmfgraph*}
	 } \\
	 &+\parbox{41mm}%36
	 {
	   \begin{fmfgraph*}(40,18)
	     \fmfleft{l}
	     \fmfright{r}
	     \fmf{plain_arrow}{r,c1}
	     \fmf{plain_arrow,tension=-0.2}{c1,c2}
	     \fmf{plain_arrow}{c2,l}  
	     \fmf{photon,right}{c1,c2}
	     \fmfv{decor.shape=square,decor.filled=empty,decor.size=4mm,label=$\sigma$,label.dist=-1mm}{c2}  
	     \fmfdot{c1}
	   \end{fmfgraph*}
	 } \\
	 &+\parbox{41mm}%36
	 {
	   \begin{fmfgraph*}(40,18)
	     \fmfleft{l}
	     \fmfright{r}
	     \fmf{plain_arrow}{r,c2}
	     \fmf{plain_arrow,tension=-0.2}{c2,c1}
	     \fmf{plain_arrow}{c1,l}  
	     \fmf{photon,right}{c2,c1}
	     \fmfv{decor.shape=square,decor.filled=empty,decor.size=4mm,label=$\sigma$,label.dist=-1mm}{c2}  
	     \fmfdot{c1}
	   \end{fmfgraph*}
	 } \\
	 &+\parbox{41mm}
	 {
	   \begin{fmfgraph*}(40,18)
	     \fmfleft{l}
	     \fmfright{r}
	     \fmf{plain_arrow}{r,b1}
	     \fmf{plain_arrow,tension=-1}{b1,c}
	     \fmf{plain_arrow,tension=-1}{c,b2}
	     \fmf{plain_arrow}{b2,l}
	     \fmfv{decor.shape=square,decor.filled=empty,decor.size=4mm,label=$\sigma$,label.dist=-1mm}{c}  
	     \fmf{photon,left}{b2,b1}
	     \fmfdot{b2,b1}
	   \end{fmfgraph*}
	 } \\
	 &+\parbox{41mm}
	 {
	   \begin{fmfgraph*}(40,18)
	     \fmfleft{l}
	     \fmfright{r}
	     \fmf{plain_arrow}{r,c1}
	     \fmf{dbl_plain_arrow,tension=-0.2}{c1,c2}
	     \fmf{plain_arrow}{c2,c3}
	     \fmf{plain_arrow}{c3,l}
	     \fmfv{decor.shape=square,decor.filled=empty,decor.size=4mm,label=$\sigma$,label.dist=-1mm}{c3}  
	     \fmf{dashes_arrow,left}{c2,c1}
	   \end{fmfgraph*}
	 } \\
	 &+\parbox{41mm}
	 {
	   \begin{fmfgraph*}(40,18)
	     \fmfleft{l}
	     \fmfright{r}
	     \fmf{plain_arrow}{r,c1}
	     \fmf{plain_arrow}{c1,c2}
	     \fmf{dbl_plain_arrow,tension=-0.2}{c2,c3}
	     \fmf{plain_arrow}{c3,l}
	     \fmfv{decor.shape=square,decor.filled=empty,decor.size=4mm,label=$\sigma$,label.dist=-1mm}{c1}  
	     \fmf{dashes_arrow,left}{c3,c2}
	   \end{fmfgraph*}
	 } \\
	 &+\parbox{41mm}
	 {
	   \begin{fmfgraph*}(40,18)
	     \fmfleft{l}
	     \fmfright{r}
	     \fmf{plain_arrow}{r,b1}
	     \fmf{dbl_plain_arrow,tension=-1}{b1,c}
	     \fmf{dbl_plain_arrow,tension=-1}{c,b2}
	     \fmf{plain_arrow}{b2,l}
	     \fmfv{decor.shape=circle,decor.filled=empty,decor.size=4mm,label=$\sigma$,label.dist=-1mm}{c}  
	     \fmf{dashes_arrow,left}{b2,b1}
	   \end{fmfgraph*}
	 }.
\end{split}
\>
We obtain\\
\< \label{EQ:phi-phi}
\begin{split} 
\left< \Phi^a \bar{\Phi}^a\right>(k)
&=-i\slashed{k}^{-1} \biggl(1+\frac{2}{\pi^2N}\log(|k|/\mu)\biggr) \\
&+ i\slashed{k}^{-1}\slashed{k}_{\sigma}\slashed{k}^{-1} \biggl(\Delta_{\sigma}+\frac{\log(|k|/\mu)}{15\pi^2N}|\epsilon_{\sigma\mu\nu}|\  F_{\sigma;\mu\nu}\biggr),
\end{split}
\>
where  $F_{\sigma;\mu\nu}$ is given by
\<
F_{\sigma;\mu\nu} = 104 \Delta_{\sigma}-32\Delta_{\mu}-32\Delta_{\nu} + 2 \delta_{\sigma} + 4\delta_{\mu}+4\delta_{\nu} .
\>
From Eq.~(\ref{EQ:phi-phi}), we obtain the scaling dimension of the $\Phi$ field in Feynman gauge :
\<
\Delta_{\Phi} = 1+\frac{1}{\pi^2 N}.
\>
Then we use the Callan-Symanzik equation machinery by scaling the momentum $k\rightarrow e^{-\ell}k$ and applying 
\begin{align}
\label{EQ:CS_Lorentz_1}  \biggl(-\frac{\partial}{\partial\ell}+\left(3-2\Delta_{\Psi}\right)+ \frac{d C_{m}}{d\ell}\frac{\partial}{\partial C_m} \biggr)\left<\Psi^a \bar{\Psi}^a\right>&=0, \\
\label{EQ:CS_Lorentz_2}  \biggl(-\frac{\partial}{\partial\ell}+\left(3-2\Delta_{\Phi}\right)+ \frac{d C_{m}}{d\ell}\frac{\partial}{\partial C_m} \biggr) \left< \Phi^a \bar{\Phi}^a \right>&=0 , 
\end{align}
where the set $C$ includes the coupling constants that flow. For the  perturbation under consideration it has 6 elements:
\begin{equation}\nonumber
C = \{\Delta_0,\Delta_1,\Delta_2,\delta_0,\delta_1,\delta_2\}.
\end{equation}
%\end{widetext}
%
Eqs.~(\ref{EQ:CS_Lorentz_1}), and~(\ref{EQ:CS_Lorentz_2}) will lead to the RG flow :
\begin{equation}%\begin{eqnarray}
\frac{\partial}{\partial \ell}
\begin{bmatrix}
\Delta_0\\
\Delta_1\\
\Delta_2\\
\delta_0\\
\delta_1\\
\delta_2\\
\end{bmatrix}
= \mathcal{D}
\begin{bmatrix}
\Delta_0\\
\Delta_1\\
\Delta_2\\
\delta_0\\
\delta_1\\
\delta_2\\
\end{bmatrix}, 
\end{equation}
where the matrix anomalous dimension $\mathcal{D}$ is given by
\begin{equation}
\mathcal{D} =
\frac{1}{15\pi^2 N}
\begin{bmatrix}
-74 & 32 & 32 & -2 & -4 & -4 \\
32 & -74 & 32 & -4 & -2 & -4 \\
32 & 32 & -74 & -4 & -4 & -2\\
 -2 & -4 & -4 & -10 & 0 & 0 \\
 -4 & -2 & -4 & 0 & -10 & 0\\
 -4 & -4 & -2 & 0 & 0 & -10\\
\end{bmatrix} .
\end{equation}%\end{eqnarray} %
All the eigenvalues of this matrix anomalous dimension is nonpositive. There is one zero eigenvalue with the corresponding eigenvector
\begin{eqnarray}
  \begin{bmatrix}
    &\Delta&\\
    &\Delta&\\
    &\Delta&\\
    -&\Delta&\\
    -&\Delta&\\
    -&\Delta&\\
  \end{bmatrix} . \nonumber
\end{eqnarray}
The above shift can be absorbed by scaling $\Psi$ and $\Phi$ fields separately. The scaling is \textit{delicate}, since in our RG scheme we have fixed the coefficient of $\left( b\bar{\Psi}^a \Phi^a + c.c. \right)$ to be 1. So the scaling of $\Psi$ and $\Phi$  has to be accompanied by an appropriate scaling of $b$. But this - in general - will move us away from the critical point. For the above perturbation though, the scaling factor for $b$ is 1: %
\begin{equation} 
\nonumber \frac{1}{\sqrt{1-\Delta}\sqrt{1+\Delta}}=1+\mathcal{O}(\Delta^2).
\end{equation}
So the above perturbation remains \textit{exactly marginal} to all orders in $1/N$.

It is interesting to note that the uniform shift 
\begin{equation} \nonumber
\begin{bmatrix}
  \Delta\\
  \Delta\\
  \Delta\\
  \Delta\\
  \Delta\\
  \Delta\\
\end{bmatrix}
\end{equation}
is an eigenvector with a negative eigenvalue. Why should this perturbation flow? The absorbtion of the uniform perturbation in $\Psi$ and $\Phi$ has to be
accompanied by scaling the $b$ field by 
\begin{equation} 
\frac{1}{\sqrt{1+\Delta}\sqrt{1+\Delta}}=1-\Delta+\mathcal{O}(\Delta^2).
\end{equation} 
This scaling of the $b$ field will move us away from the critical point.
%********************************************************************************
\section{Staggered Spin Correlation \label{APP:Staggered Spin}}
%********************************************************************************
In this appendix we will find the algebraic decay exponent of $\left<\bm{M}(x) \cdot \bm{M}(x')\right>$ at KI--ASL fixed point. We do this by obtaining the staggered spin $\bm{M}$ scaling dimension. It suffices to find the scaling dimension of $M^{+}$, since the decay exponent of $\left<\bm{M}(x) \cdot \bm{M}(x')\right>$ and $\left<M^{+}(x) M^{-}(x')\right>$ is the same. The first step is to translate $M^{+}$ to the field theory language with the dictionary provided by Eq.~\ref{EQ:L_E2}. The expanded form is given by
\begin{eqnarray}
  %
  %\Psi &=&
  %\begin{bmatrix}
  %  c_{A+}\\
  %  c_{B+}\\
  %  -i c_{B-} \\
  %  i c_{A-}\\
  %\end{bmatrix} \nonumber \\
  %
  %\bar{\Psi} &=&
  %\begin{bmatrix}
  %  c_{A+}\d & -c_{B+}\d & i c_{B-}\d & i c_{A-}\d \\
  %\end{bmatrix} \nonumber \\
  %
  \Phi &=&
  \begin{bmatrix}
    f_{A+}\\
    -f_{B+}\\
    -i f_{B-} \\
    -i f_{A-}\\
  \end{bmatrix} \\
  \bar{\Phi} &=&
  \begin{bmatrix}
    f_{A+}\d & f_{B+}\d & i f_{B-}\d & -i f_{A-}\d \\
  \end{bmatrix} 
\end{eqnarray}
Start with the expression for $S_A^{+}=S_A^1+iS_A^2=f_A^{1 \dagger} f_A^2$, which leads to 
\<
  S_A^{+}(x)=\bar{\Phi}^1(x)
  \begin{pmatrix}
    1&&& \\
    &0&& \\
    &&0&\\
    &&&-1\\
  \end{pmatrix}
  \Phi^2(x) .
\>
The similar translation can be done for $S_B^+$. The final result is:
\<
M^+(x) = \bar{\Phi}^1(x) \mathcal{M} \Phi^2(x) ,
\>
where $\mathcal{M}$ is 
\<
\mathcal{M} = 
\begin{pmatrix}
1 &  &  & \\
 & 1 &  &  \\
& & -1 & \\
  &  &  & -1 \\
\end{pmatrix} = \mathbf{1} \otimes \sigma_3.
\>
%\sout{Notation $\mathcal{M}$ is used since $M^-(x)$ results in the same matrix.}
The matrix $\mathcal{M}$  has the following properties:
\begin{align}
[\mathcal{M},\gamma_{\mu}] &=0, \\
\rm{Tr} \bigl(\mathcal{M} \Gamma \bigr) &= 0 ,
\end{align}
where $\Gamma$ is a generic notation representing \textit{any} number of $\gamma_{\mu}$ matrices multiplied. The 1st property is useful since $\mathcal{M}$ passes through all the $\gamma$ matrices and the final result for the propagatros $\left<\Phi_i^1 \bar{\Phi}_j^2\right>$ and $\left<\Psi_i^1 \bar{\Psi}_j^2\right>$ will have a simple matrix form $\left(\slashed{k}^{-1}\mathcal{M}\slashed{k}^{-1}\right)_{ij}$. The second property causes any diagram, with matrix  $\mathcal{M}$ left in a loop,  to vanish. 

The scaling dimension of $M^+$ is obtained by looking at its relevant flow near the fixed point. Due to mixing we perturb the Lagrangian density by % 
\begin{equation}
\delta \mathcal{L}_E= U_+\ \bar{\Phi}^1(x) \mathcal{M} \Phi^2(x) + u_+
\ \bar{\Psi}^1(x) \mathcal{M} \Psi^2(x),
\end{equation} %
and study the Callan-Symanzik equation for the propagators $\left<\Phi^1 \bar{\Phi}^2\right>$ and $\left<\Psi^1 \bar{\Psi}^2\right>$.

The vertices corresponding to these perturbations are represented by
\<
\begin{split}
  \parbox{33mm}{
    \begin{fmfgraph*}(30,18)
      \fmfstraight
      \fmfleft{i}
      \fmfright{o}
      \fmf{plain_arrow,label=$2$}{o,c}
      \fmf{plain_arrow,label=$1$}{c,i}
      \fmfv{decor.shape=diamond,decor.filled=1,decor.size=4mm}{c}
    \end{fmfgraph*}
  } &= U_+\ \mathcal{M} \\
  \parbox{33mm}{
    \begin{fmfgraph*}(30,18)
      \fmfstraight
      \fmfleft{i}
      \fmfright{o}
      \fmf{dbl_plain_arrow,label=$2$}{o,c}
      \fmf{dbl_plain_arrow,label=$1$}{c,i}
      \fmfv{decor.shape=diamond,decor.filled=0,decor.size=4mm}{c}
    \end{fmfgraph*}
  } &= u_+\ \mathcal{M}
\end{split}
\>
The diagrammatic expression for propagators $\left<\Phi^1 \bar{\Phi}^2\right>$ and $\left<\Psi^1 \bar{\Psi}^2\right>$ is then given by:
\<
\begin{split}
  \left<\Psi^1 \bar{\Psi}^2\right> &= 
  \parbox{51mm}%28
	 {
	   \begin{fmfgraph*}(50,18)
	     \fmfstraight
	     \fmfleft{i}
	     \fmfright{o}
	     \fmf{dbl_plain_arrow}{o,c}
	     \fmf{dbl_plain_arrow}{c,i}
	     \fmfv{decor.shape=diamond,decor.filled=0,decor.size=4mm}{c}
	   \end{fmfgraph*}
	 } \\
	 &+\parbox{51mm}
	 {
	   \begin{fmfgraph*}(50,18)
	     \fmfleft{l}
	     \fmfright{r}
	     \fmf{dbl_plain_arrow}{r,c1}
	     \fmf{plain_arrow,tension=-0.2}{c1,c2}
	     \fmf{dbl_plain_arrow}{c2,c3}
	     \fmf{dbl_plain_arrow}{c3,l}
	     \fmfv{decor.shape=diamond,decor.filled=0,decor.size=4mm}{c3}
	     %  \fmfv{decor.shape=circle,decor.filled=empty,decor.size=4mm,label=$\sigma$,label.dist=-1mm}{c3}  
	     \fmf{dashes_arrow,right}{c1,c2}
	   \end{fmfgraph*}
	 } \\
	 &+\parbox{51mm}
	 {
	   \begin{fmfgraph*}(50,18)
	     \fmfleft{l}
	     \fmfright{r}
	     \fmf{dbl_plain_arrow}{r,c1}
	     \fmf{dbl_plain_arrow}{c1,c2}
	     \fmf{plain_arrow,tension=-0.2}{c2,c3}
	     \fmf{dbl_plain_arrow}{c3,l}
	     \fmfv{decor.shape=diamond,decor.filled=0,decor.size=4mm}{c1}
	     %    \fmfv{decor.shape=circle,decor.filled=empty,decor.size=4mm,label=$\sigma$,label.dist=-1mm}{c1}  
	     \fmf{dashes_arrow,right}{c2,c3}
	   \end{fmfgraph*}
	 } \\
	 &+\parbox{51mm}
	 {
	   \begin{fmfgraph*}(50,18)
	     \fmfleft{l}
	     \fmfright{r}
	     \fmf{dbl_plain_arrow}{r,b1}
	     \fmf{plain_arrow,tension=-1}{b1,c}
	     \fmf{plain_arrow,tension=-1}{c,b2}
	     \fmf{dbl_plain_arrow}{b2,l}
	     \fmfv{decor.shape=diamond,decor.filled=1,decor.size=4mm}{c}
	     %    \fmfv{decor.shape=square,decor.filled=empty,decor.size=4mm,label=$\sigma$,label.dist=-1mm}{c}  
	     \fmf{dashes_arrow,right=0.7}{b1,b2}
	   \end{fmfgraph*}
	 } ,
\end{split}
\>
\<
\begin{split}
	 \left<\Phi^1 \bar{\Phi}^2\right> &=
	 \parbox{51mm}%28
		{
		  \begin{fmfgraph*}(50,18)
		    \fmfstraight
		    \fmfleft{i}
		    \fmfright{o}
		    \fmf{plain_arrow}{o,c}
		    \fmf{plain_arrow}{c,i}
		    \fmfv{decor.shape=diamond,decor.filled=1,decor.size=4mm}{c}
		  \end{fmfgraph*}
		} \\
		&+\parbox{51mm}
		{
		  \begin{fmfgraph*}(50,18)
		    \fmfleft{l}
		    \fmfright{r}
		    \fmf{plain_arrow}{r,c1}
		    \fmf{plain_arrow,tension=-0.2}{c1,c2}
		    \fmf{plain_arrow}{c2,c3}
		    \fmf{plain_arrow}{c3,l}
		    \fmfv{decor.shape=diamond,decor.filled=1,decor.size=4mm}{c3}  
		    \fmf{photon,left}{c2,c1}
		    \fmfdot{c1,c2}
		  \end{fmfgraph*}
		} \\
		&+\parbox{51mm}
		{
		  \begin{fmfgraph*}(50,18)
		    \fmfleft{l}
		    \fmfright{r}
		    \fmf{plain_arrow}{r,c1}
		    \fmf{plain_arrow}{c1,c2}
		    \fmf{plain_arrow,tension=-0.2}{c2,c3}
		    \fmf{plain_arrow}{c3,l}
		    \fmfv{decor.shape=diamond,decor.filled=1,decor.size=4mm}{c1}  
		    \fmf{photon,left}{c3,c2}
		    \fmfdot{c3,c2}
		  \end{fmfgraph*}
		} \\
		&+\parbox{51mm}
		{
		  \begin{fmfgraph*}(50,18)
		    \fmfleft{l}
		    \fmfright{r}
		    \fmf{plain_arrow}{r,b1}
		    \fmf{plain_arrow,tension=-1}{b1,c}
		    \fmf{plain_arrow,tension=-1}{c,b2}
		    \fmf{plain_arrow}{b2,l}
		    \fmfv{decor.shape=diamond,decor.filled=1,decor.size=4mm}{c}  
		    \fmf{photon,left=0.7}{b2,b1}
		    \fmfdot{b2,b1}
		  \end{fmfgraph*}
		} \\
		&+\parbox{51mm}
		{
		  \begin{fmfgraph*}(50,18)
		    \fmfleft{l}
		    \fmfright{r}
		    \fmf{plain_arrow}{r,c1}
		    \fmf{dbl_plain_arrow,tension=-0.2}{c1,c2}
		    \fmf{plain_arrow}{c2,c3}
		    \fmf{plain_arrow}{c3,l}
		    \fmfv{decor.shape=diamond,decor.filled=1,decor.size=4mm}{c3}  
		    \fmf{dashes_arrow,left}{c2,c1}
		  \end{fmfgraph*}
		} \\
		&+\parbox{51mm}
		{
		  \begin{fmfgraph*}(50,18)
		    \fmfleft{l}
		    \fmfright{r}
		    \fmf{plain_arrow}{r,c1}
		    \fmf{plain_arrow}{c1,c2}
		    \fmf{dbl_plain_arrow,tension=-0.2}{c2,c3}
		    \fmf{plain_arrow}{c3,l}
		    \fmfv{decor.shape=diamond,decor.filled=1,decor.size=4mm}{c1}  
		    \fmf{dashes_arrow,left}{c3,c2}
		  \end{fmfgraph*}
		} \\
		&+\parbox{51mm}
		{
		  \begin{fmfgraph*}(50,18)
		    \fmfleft{l}
		    \fmfright{r}
		    \fmf{plain_arrow}{r,b1}
		    \fmf{dbl_plain_arrow,tension=-1}{b1,c}
		    \fmf{dbl_plain_arrow,tension=-1}{c,b2}
		    \fmf{plain_arrow}{b2,l}
		    \fmfv{decor.shape=diamond,decor.filled=0,decor.size=4mm}{c}
		    \fmf{dashes_arrow,left=0.7}{b2,b1}
		  \end{fmfgraph*}
		} .
\end{split}
\>
\newpage
The final answers in the Feynman gauge are as follows:
\begin{widetext}
%\begin{eqnarray}
%  \left<\Psi^1 \bar{\Psi}^2\right>(k) =-\slashed{k}^{-1}\mathcal{M}\slashed{k}^{-1} &\Biggl(& u_+ \left(1+\frac{4}{3\pi^2 N} \log(|k|/\mu)\right) \nonumber \\
%  &\ & + \ \frac{2U_+ }{\pi^2 N} \log(|k|/\mu)\Biggr) , \\
%  \left<\Phi^1 \bar{\Phi}^2\right>(k) =  -\slashed{k}^{-1}\mathcal{M}\slashed{k}^{-1} &\Biggl(& U_+ \left(1-\frac{8}{\pi^2 N} \log(|k|/\mu)\right) \nonumber \\
%  &\ & + \ \frac{2u_+ }{\pi^2 N} \log(|k|/\mu)\Biggr). 
%\end{eqnarray}

\begin{eqnarray}
  \left<\Psi^1 \bar{\Psi}^2\right>(k) =-\slashed{k}^{-1}\mathcal{M}\slashed{k}^{-1} &\Biggl(& u_+ \left(1+\frac{4}{3\pi^2 N} \log(|k|/\mu)\right) + \ \frac{2U_+ }{\pi^2 N} \log(|k|/\mu)\Biggr) , \\
  \left<\Phi^1 \bar{\Phi}^2\right>(k) =  -\slashed{k}^{-1}\mathcal{M}\slashed{k}^{-1} &\Biggl(& U_+ \left(1-\frac{8}{\pi^2 N} \log(|k|/\mu)\right)+ \ \frac{2u_+ }{\pi^2 N} \log(|k|/\mu)\Biggr). 
\end{eqnarray}

\end{widetext}

Then we use the Callan-Symanzik equation by scaling the momentum $k\rightarrow e^{-\ell}k$ and applying
\<
\begin{split}  \label{EQ:CS_SSC}
  \biggl(-\frac{\partial}{\partial\ell}+\left(3-2\Delta_{\Psi}\right)+ \frac{d C_{m}}{d\ell}\frac{\partial}{\partial C_m} \biggr)\left<\Psi^1 \bar{\Psi}^2\right>&=0 \\
  \biggl(-\frac{\partial}{\partial\ell}+\left(3-2\Delta_{\Phi}\right)+ \frac{d C_{m}}{d\ell}\frac{\partial}{\partial C_m} \biggr) \left< \Phi^1 \bar{\Phi}^2 \right>&=0 
\end{split}
\>
The set $C$ includes the coupling constants that flow. For the  perturbation under consideration it has two elements:
\begin{equation}\nonumber
C = \{U_{+},u_{+}\}.
\end{equation}
The above Callan-Symanzik equations leads to the following RG flow:
\begin{equation}
\frac{\partial}{\partial \ell}
\begin{bmatrix}
U_+\\
\\
u_+\\
\end{bmatrix}
=
\begin{bmatrix}
1+\frac{10}{\pi^2 N} & \frac{-2}{\pi^2 N}  \\
 & \\
\frac{-2}{\pi^2 N} & 1-\frac{2}{3 \pi^2 N} \\
\end{bmatrix}
\begin{bmatrix}
U_+\\
\\
u_+\\
\end{bmatrix}.
\end{equation}

\end{fmffile}

%\newpage
%\vspace{6cm}
%\bibliography{khh}

\end{document}